\documentclass[aps,prb,twocolumn,floatfix,amsmath,amssymb,footinbib,showpacs]{revtex4}

\usepackage{graphicx}
\usepackage{textgreek}

\begin{document}
\title{Non-equilibrium transport through a Josephson quantum dot}
\author{J.F. Rentrop}
\author{S.G. Jakobs}
\author{V. Meden}
\affiliation{\small Institut f\"ur Theorie der Statistischen Physik, RWTH Aachen University and\\ \small JARA---Fundamentals of Future Information Technology, 52062 Aachen, Germany}
\date{\vspace{-5ex}}

\begin{abstract}
We study the electronic current through a quantum dot coupled to two superconducting leads 
which is driven by either a voltage $V$ or temperature $\Delta T$ bias. Finite biases beyond 
the linear response regime are considered. The local two-particle interaction $U$ on the dot is 
treated using an approximation scheme within the functional renormalization 
group approach set up in Keldysh-Nambu-space with $U$ being the small parameter. For $V>0$ we 
compare our renormalization group enhanced results for the dc-component of the current to earlier weak coupling approaches such as 
the Hartree-Fock approximation and second order perturbation theory in $U$. We show that in 
parameter regimes in which finite bias driven multiple Andreev reflections prevail small 
$|U|$ approaches become unreliable for interactions of appreciable strength. In the complementary 
regime the convergence of the current with respect to numerical parameters becomes an issue---but can eventually be achieved---and interaction effects turn out to be 
smaller then expected based on earlier results. For $\Delta T>0$ we find a surprising increase 
of the current as a function of the superconducting phase difference in the regime which at $T=0$ 
becomes the $\pi$ (doublet) phase.  
\end{abstract}
\pacs{73.21.La, 73.23.-b, 73.63.-b, 74.45.+c, 74.50.+r}

\maketitle

\section{Introduction}
\label{sec:Introduction}
Mesoscopic systems of two BCS superconductors coupled via a quantum dot show rich physics. This holds even if the dots one-particle level spacing is much larger than the 
reservoir-dot coupling $\Gamma$, temperatures $T_L$ and $T_R$ of the left ($L$) and right ($R$)
reservoirs, and bias voltage $V$ applied across the dot. In this case, studying the simplified model 
of a quantum dot with a single, spin-degenerate level becomes meaningful; we focus on this situation
and refer to it as the Josephson quantum dot.  
The proximity effect induces a superconducting gap on the dot and, for superconducting
phase differences $\phi$ different from integer multiples of $\pi$, a Josephson current runs through the 
dot even in equilibrium $V=\Delta T = T_L - T_R =0$.\cite{Mar11} The case of finite bias 
voltages---but still $\Delta T=0$---was intensively studied over the past 
years\cite{Yey97,Kan98,Joh99,Avi01,Yey03,Liu04,Zaz06,Del08,And11,Bui02,Bui03,Eic07,San07,Hil12} 
and allows for multiple Andreev reflections (MAR) as well as the AC Josepshon effect 
both leaving their signatures in the electronic current. These effects also appear in 
tunnel-contacted superconductors (to be contrasted to quantum dot contacted ones considered here). 
In the case in which the dots charging energy or equivalently the local Coulomb repulsion $U$ vanishes, 
the Josephson current, MAR as well as the AC Josephson effect are well understood and all parameter 
dependencies can straightforwardly be computed.\cite{Mar11,Yey97,Joh99,Bui03,Eic07,Hil12}

In MAR, an electron passing the dot gains an energy of $V>0$ and is then reflected at the superconductor 
as a hole which once more aquires an energy of $V$ as it passes the dot. Then the hole is reflected 
and the process starts again with an electron. Only after having gained an energy of $(2n+1)V$ by 
multiple reflections, the electron has gathered enough energy to overcome the superconducting gap of full 
width $2\Delta$. This makes it plausible that at each voltage fulfilling $2\Delta/V=2n+1, n \in \mathbb{N}_0$, 
a new transport channel opens up significantly contributing to the current $I$ and thus explaining the 
characteristic features of an $I(V)$ curve at these voltages. Interactions will have a great effect at the 
MAR points; the opening of a MAR channel implies a highly fluctuating occupation 
of the dot, susceptible to electronic correlations. 

For sizable $U$ novel effects appear. For example, the interplay between the superconducting gap 
$\Delta$ and the Coulomb 
energy triggers for $V=0$ and $T=0$ a first order level-crossing quantum phase transition between 
a singlet and a doublet phase.\cite{Mar11} 
The phase boundary can be crossed by variation of the two-particle interaction $U$, the superconducting
phase difference $\phi$, and the dot level energy. 
In the context of bulk superconductors with magnetic impurities 
this phase transition was revealed and qualitatively understood already decades 
ago.\cite{Sod67,Shi69,Zit70,Mue70,Zit70a,Mue71,Shi73,Mat76,Spi91} Recent 
theoretical\cite{Gla89,Roz99,Vec03,Ogu04,Sia04,Cho04,Sia05,Cho05,Nov05,Tan07,Bau07,Kar08,Men09,Lui10,Dro12}  
and experimental progress\cite{Tan97,Kas99,Dam06,Cle06,Joe07,Eic09,Mau12} added a quantitative 
understanding for mesoscopic 
systems including the effect of finite but equal  temperatures ($\Delta T=0$).\cite{Sia04,Kar08,Lui10}
It was even possible 
to achieve a satisfying agreement between experimental data for the critical current obtained with a carbon nanotube 
as the quantum dot\cite{Joe07} and model calculations.\cite{Lui12} Depending 
on the details of the experimental setup the Kondo effect\cite{Hewson} becomes important 
and interesting physics result out of an interplay of superconducting and Kondo 
correlations.\cite{Sod67,Shi69,Zit70,Mue70,Zit70a,Mue71,Shi73,Mat76,Ogu04,Sia04,Cho04,Sia05,Cho05,Tan07,Bau07,Kar08,Lui10} As later on we will mainly 
consider regimes in which the Kondo effect does not develop, we do not describe the details of these physics.   
Beautiful spectroscopic data\cite{Pil10,Dea10,Fra11,Lee12,Bau13} triggered much current effort to study the parameter dependencies of the energy of the Andreev bound states (ABS)\cite{Mar11,Bau07,Men09} which mainly carry the current.

For $V>0$, the interaction effects are by far less well understood. Model calculations\cite{Kan98,Avi01,Yey03,Liu04,Zaz06,Del08,And11,Bui03,Eic07,San07} allowed to understand certain aspects of e.g. 
the experimentally observed MAR features appearing in the current but a comprehensive picture did 
up to now not emerge. In fact, treating the combined problem of superconductivity, local two-particle 
interaction, and non-equilibrium provides a sizeable theoretical challenge that has not yet been overcome. In this paper, we study to what extent the functional renormalization group (RG) can contribute to clarifying this situation. As experiments are often performed 
in parameter regimes in which charge fluctuations are not fully suppressed (that is spin fluctuations do not 
prevail), one cannot resort to exclusively studying the Kondo model\cite{Hewson} with superconducting leads. 
Therefore we here consider the superconducting extension of the single-impurity Anderson model (SIAM).\cite{Hewson}
         
To study the $V>0$ dc-current through a Josephson quantum dot we use an approximation method 
which is derived from a general functional RG technique\cite{Met12} set up in 
Keldysh-Nambu-space with $U>0$ being the small parameter. The functional RG approach to mesoscopic transport\cite{Met12} 
was earlier generalized to study non-equilibrium setups with metallic leads in the steady state\cite{Gez07,Jak07,Jak10a,Jak10c,Kar10} as well as ground state properties of quantum dots with 
superconducting leads such as the Josephson current.\cite{Kar08,Eic09,Kar10,Kar11} We here combine 
these two extensions and thus further advance the method. We show that in parameter regimes in which 
finite bias driven multiple Andreev reflections prevail ($\Gamma \lesssim \Delta$) even 
RG enhanced small $U$ approaches become unreliable for interactions of appreciable strength; within our 
approximate approach we can perform a consistency check which allows us to identify values of $U$ for which 
it becomes uncontrolled (at fixed other parameters). The functional RG results turn out to be similar to the ones obtained 
using the restricted self-consistent Hartree-Fock (SCHF) approximation and second order perturbation 
theory;\cite{Del08} for a SCHF approach 
relying on additional approximations see Ref.\ \onlinecite{Avi01}. In the complementary regime 
with small MAR features in the current ($\Gamma > \Delta$), the convergence of the current with respect to parameters appearing 
in the numerical solution of the weak coupling equations becomes an issue; carefully ensuring numerical convergence, we show that interaction effects 
turn out to be significantly smaller then expected based on earlier results. 

Temperature gradients across nano- and mesoscopic systems are difficult to be realized experimentally. Thus, non-equilibrium currents across the Josephson quantum dot driven by such were so far not in the 
focus of theoretical studies and we present some first calculations in this regime. We show that for $\Delta T>0$ (but $V=0$) a surprising increase of the 
current as a function of the superconducting phase difference $\phi$ appears in the regime which at 
$T=0$ becomes the doublet phase. We hope that this result will motivate further theoretical research on 
Josephson quantum dots in the $\Delta T>0$ non-equilibrium steady state and ultimately also experiments in this situation.
    
The paper is organized as follows: In Sect.\ \ref{sec:model}, the Hamiltonian is presented. Also, the 
required single-particle Green functions are introduced as well as two special Fourier transformations described
in detail in App.\ \ref{sec:FT}. In Sect.\ \ref{sec:flow_eq}, the functional RG flow equations are 
motivated and presented explicitly, while some more general remarks about functional RG can be found in App.\ \ref{sec:Details_fRG}. 
The equations showing how to compute the current after the RG flow and how to perform the self-consistency loop in SCHF are given in Sect.\ \ref{sec:Formula_current}. Details on the numerical solution of the RG flow equations and results are discussed in Sect.\ \ref{sec:num_results}. 
Our main findings are summarized in Sect.\ \ref{sec:conclusion}.

\section{Model and Keldysh Green functions}
\label{sec:model}
In Nambu form the SIAM with superconducting leads is given by the Hamiltonian 
\begin{equation}
H=H_{\textrm{dot}}+\sum_{s=\textrm{L,R}=\mp} H_s^{\textrm{coup}}+H_s^{\textrm{lead}}
\end{equation}
with the dot part
\begin{equation} 
H_{\textrm{dot}} = \varphi^\dagger \widetilde{\epsilon}_3 \varphi - 
U \varphi_0^\dagger \varphi_1^\dagger \varphi_1 \varphi_0 ,
\end{equation}
where
\begin{equation}
\widetilde{\epsilon}_3=\left( \begin{array}{cc} \epsilon_\uparrow+U/2 & 0 \\ 0 & -(\epsilon_\downarrow-U/2) 
\end{array} \right) . \end{equation} 
The BCS leads are modeled as 
\begin{equation}
H_s^{\textrm{lead}}= \sum_{k} \psi_{s,k}^\dagger \left( \begin{array}{cc} \epsilon_{s,k} 
& -\Delta_s e^{i \phi_s} \\ -\Delta_s e^{-i \phi_s} & -\epsilon_{s,-k} \end{array} \right) \psi_{s,k} \end{equation}
and the lead-dot coupling is 
\begin{align} H_s^{\textrm{coup}} =\hphantom{+}& \psi_s^\dagger \sigma_z 
\left( -t_s e^{s i \sigma_z t V/2} \right) \varphi \\ \notag +& \varphi^\dagger \sigma_z 
\left( -t_s e^{-s i \sigma_z t V/2} \right) \psi_s 
\end{align}
Here,  $\varphi^\dagger=(d_\uparrow^\dagger,d_\downarrow)$ is the Nambu dot creation operator where 
$d_\sigma^{(\dag)}$ denote the electronic dot ladder operators. Similarly, $\psi^\dagger_{s,k}=(c_{\uparrow s,k}^\dagger,c_{\downarrow s,-k})$ 
for the leads (where $k$ denotes the momentum). The Nambu index $q=0,1$ replaces the spin index 
$\sigma=\uparrow,\downarrow=\pm$. Furthermore, $\psi_s=\sum_k \psi_{s,k} /\sqrt{N}$ is the Nambu annihilation 
operator at the end of the lead ($N$ is the number of $k$ modes). 
The one-particle energies $\epsilon_\sigma=V_\textrm{g}-\sigma B$ depend on a possible local 
Zeeman field $B$ and can be varied by tuning a gate voltage $V_\textrm{g}$; $V_\textrm{g}=0$ corresponds to 
particle-hole symmetry. The Hamiltonian is written in a particular electro-magnetic gauge
which renders the dot-lead coupling part explicitly time-dependent; the bias voltage $V$ enters via a time-dependent
phase factor.\cite{Rog74} 
We assumed that the tunnel amplitudes $t_\textrm{L/R}$ are independent of spin and real valued. Phase factors in the $t_{L/R}$ could be absorbed into the superconducting complex phases $e^{i\phi_\textrm{L/R}}$.
 
Within our functional RG approach \cite{Met12,Jak10c} the single-particle irreducible 
vertex functions are computed. Observables of interest such as the current can be 
determined from these vertex functions (for details, see Sect.\ \ref{sec:Formula_current}). Basic elements of the functional RG are the dot Green functions. 
With respect to the Nambu structure ($q=0,1$; see above) the retarded and Keldysh ones are defined as
\begin{equation}
G^\textrm{R}_{q q^\prime}(t,t^\prime)=-i \theta(t-t^\prime)
\left\langle\left[\varphi_q(t),\varphi^\dagger_{q^\prime}(t^\prime)\right]_+\right\rangle ,
\end{equation}
\begin{equation}
G^\textrm{K}_{q q^\prime}(t,t^\prime)=-i \left\langle\left[\varphi_q(t),\varphi^\dagger_{q^\prime}(t^\prime)\right]_-\right\rangle ,
\end{equation}
with the commutator $[ \ldots , \ldots ]_-$ and anti-commutator $[ \ldots , \ldots ]_+$. For the advanced component, it holds $G^\textrm{A}_{q q^\prime}(t,t^\prime)=\left( G^\textrm{R}_{q^\prime q}(t^\prime,t) \right)^\ast$. Sometimes, the retarded, Keldysh and advanced components are arranged in a matrix structure, which here is referenced by indices $\alpha,\alpha^\prime$, where the following mapping holds: $G^\textrm{R} = G^{10}$, $G^\textrm{K} = G^{11}$ and $G^\textrm{A} = G^{01}$.

From now on, the $V>0$ and $V=0$ case will be discussed separately as they imply a fundamentally different Hamiltonian either featuring a time-dependence or not. The discussion of the $V>0$ case will always precede the one of the $V=0$ case.

\subsection{Keldysh Green functions for $\mathbf{V>0}$}
For $V>0$, the time-dependence within the Hamiltonian is periodic: $H(t)=H(t+T)$ where $T=2\pi/V$. This implies a global periodicity for the Green and vertex functions, e.g. $G^{\alpha \alpha^\prime}_{q q^\prime}(t_1,t_1^\prime)=G^{\alpha \alpha^\prime}_{q q^\prime}(t_1+T,t_1^\prime+T)$. Two Fourier transforms (FT) are used in this work to exploit this. Combining the two times linearly to a centered and a relative time, it is apparent that due to the global periodicity a discrete Fourier index is sufficient to transform the centered time, whereas a continuous Fourier frequency on the entire real axis is needed to transform the relative time. This idea is called single-indexed FT (siFT). It turns out that for single-particle functions an equivalent transform can be formulated that employs two discrete Fourier indices and one real Fourier frequency within the interval $[-V/2,V/2)$ (called double-indexed FT---diFT).\cite{Del08,Arn87,Mar99} The siFT has the advantage that it can be generalized to the many-particle case without losing its property of exploiting the global periodicity. The diFT has the advantage that the inversion of a single-particle quantity corresponds to a matrix inversion. Details about siFT and diFT are given in App.\ \ref{sec:FT}.

The retarded component of the inverse free propgator in diFT reads ($\omega_n=\omega+nV$, $\omega \in [-V/2,V/2)$):
\begin{equation} \label{eq:free_HF_propagator}
\left(G_{\textrm{free}}^{-1}\right)^{\textrm{R}}(\omega)_{n^\prime n}
=  \delta_{n^\prime n} \left( \begin{array}{cc} \omega_n - \epsilon_\uparrow + i\eta & 0 \\ 0 & \omega_n + \epsilon_\downarrow +i\eta \end{array} \right) 
\end{equation}
Note that we redefined the inverse propagator and accordingly the self-energy by subtracting the Hartree shift $-U/2$.
The Keldysh component is given by 
\begin{equation}
\left(G_{\textrm{free}}^{-1}\right)^{\textrm{K}}(\omega)_{n^\prime n}
=  2i\eta \delta_{n^\prime,n}[1-2f_\textrm{dot}(\omega_n)]\left(\begin{array}{cc}1 
& 0 \\ 0 & 1\end{array}\right) . 
\end{equation}
It is proportional to the positive real number $\eta$ which must be sent to zero at the end of all computations. It turns out that for $V > 0$, $\eta$ can be sent to zero from the outset which simplifies calculations. Thus, the precise choice for $f_\textrm{dot}$ is not critical for $V>0$. The physical meaning of $\eta$ and $f_\textrm{dot}$ (and its choice) is discussed in Sect.\ \ref{sec:Green_functions_V_0}.

The wideband limit ($\rho_0=\textrm{const.}$) is assumed for the calculation of the dot self-energy resulting from the coupling to the superconducting leads. With $\Sigma^\textrm{res,R} = \Sigma^{\textrm{res},01}$, $\Sigma^\textrm{res,K} = \Sigma^{\textrm{res},00}$ and $\Sigma^\textrm{res,A} = \Sigma^{\textrm{res},10}$, one finds (cf.\ Ref.\ \onlinecite{Del08}):
\begin{align} 
& \Sigma_{s}^{\textrm{res},0 \alpha}(\omega)_{n^\prime n}
\\ \notag = & \left( \begin{array}{cc} \! \! \delta_{n^\prime ,n} \Sigma_{s,00}^{\textrm{res},0 \alpha}(\omega_n\!-\!\frac{sV}{2}) & \delta_{n^\prime-s,n} \Sigma_{s,01}^{\textrm{res},0 \alpha}(\omega_n\!+\!\frac{sV}{2}) \\ \! \! \delta_{n^\prime+s,n} \Sigma_{s,10}^{\textrm{res},0 \alpha}(\omega_n\!-\!\frac{sV}{2}) & \delta_{n^\prime,n} \Sigma_{s,11}^{\textrm{res},0 \alpha}(\omega_n\!+\!\frac{sV}{2}) \end{array} \right) ,
\end{align} 
where $\Sigma^{\textrm{res}, 0 \alpha}_{s, q^\prime q}(\omega_n \pm V/2)$ denotes the self-energy due to the leads at $V=0$ as given in Eq.\ \eqref{eq:Sigma_res_R_V_0} and \eqref{eq:Sigma_res_K_V_0}.

Certain symmetries hold for the Green and vertex functions: Complex conjugation corresponds to $K^{\alpha \alpha^\prime}_{q q^\prime}(\omega)_{n n^\prime}=-(-1)^{\alpha+\alpha^\prime}K^{\alpha^\prime \alpha}_{q^\prime q}(\omega)^\ast_{n^\prime n}$ with $K=G,\Sigma$. Complex conjugation can be used to relate the retarded and advanced components such that only the retarded one must be stored and evaluated in numerical calculations. Also, it implies that the effort of inverting a single-particle quantity reduces to inverting its retarded component; for example: $G^\textrm{R}=[(G^{-1})^\textrm{R}]^{-1}$ and $G^\textrm{K}=-G^\textrm{R}(G^{-1})^\textrm{K}(G^\textrm{R})^\dagger$.
$B=0$ corresponds to the following symmetry:  $K^{\alpha \alpha^\prime}_{q q^\prime} (\omega)_{n n^\prime}=-(-1)^{q-q^\prime}K^{\alpha^\prime \alpha}_{\bar{q}^\prime \bar q} (-\omega)_{-n^\prime,-n}$ with $\bar{q}=1-q$. It can be derived from the observation that the Hamiltonian is spin-flip invariant for $B=0$ if simultaneously $\Delta_s  \to -\Delta_s$. The symmetry can be employed to write an optimized $B=0$ code to solve the functional RG flow equations (see below) or as a numerical check of a general (arbitrary $B$) code.
Swapping particles within a many-particle function results in a minus sign.

\subsection{Keldysh Green functions for $\mathbf{V=0}$}
\label{sec:Green_functions_V_0}

For $V=0$ (remember that with $\Delta T \neq 0$ a non-equilibrium set-up can still be realized), the quantities aquire simpler structures as no explicit time-dependence in the Hamiltonian needs to be treated. The inverse free propagator is (now, $\omega \in \mathbb{R}$):
\begin{equation} \label{eq:free_HF_propagator_V_0}
\left(G_{\textrm{free}}^{-1}\right)^{\textrm{R}}(\omega)
=  \left( \begin{array}{cc} \omega - \epsilon_\uparrow + i\eta & 0 \\ 0 & \omega + \epsilon_\downarrow +i\eta \end{array} \right) 
\end{equation}
\begin{equation}
\left(G_{\textrm{free}}^{-1}\right)^{\textrm{K}}(\omega)
=  2i\eta [1-2f_\textrm{dot}(\omega)]\left(\begin{array}{cc}1 
& 0 \\ 0 & 1\end{array}\right) 
\end{equation}
For $V=0$, $\eta$ must have finite values throughout the calculations since the self-energy due to the leads (see below) does not provide a finite imaginary part within the superconducting gap. Such a positive imaginary part represents decay channels and is required to assure the emergence of a stationary state. Physically speaking, the $\eta$ can be associated with a small coupling to a (metallic) background---which can plausibly be argued to always be present. This provides a physical meaning for $f_\textrm{dot}$. It is the Fermi function of the background. The question remains what temperature $T_\textrm{dot}$ should be assigned to this background---especially in the case of $\Delta T \neq0$. It is physically reasonable to use $T_\textrm{dot}=(T_\textrm{L}+T_\textrm{R})/2$. The choice of $T_\textrm{dot}$ is indeed relevant for the numerical results; for instance, in equilibrium we found that numerical results for $T_\textrm{dot}=(T_\textrm{L}+T_\textrm{R})/2=T_\textrm{L}=T_\textrm{R}$ are closer to NRG data of Ref.\ \onlinecite{Kar08} than for $T_\textrm{dot}=0,\infty$.

Defining $\Gamma_s=\pi \rho_0 t_s^2$ (note that there is no factor of $2$), the dot self-energy resulting from the coupling to the superconducting leads is:
\begin{align}\label{eq:Sigma_res_R_V_0}
\Sigma_s^{\textrm{res,01}}(\omega)=&\Sigma_s^{\textrm{res,R}}(\omega)
\\ \notag =&- \Gamma_s W_s(\omega) \left( \begin{array}{cc} \omega & \Delta_s e^{i \phi_s} \\ \Delta_s e^{-i \phi_s} & \omega \end{array} \right)
\end{align}
\begin{align}\label{eq:Sigma_res_K_V_0}
\Sigma_s^{\textrm{res,00}}(\omega)=&\Sigma_s^\textrm{res,K}(\omega)
\\ \notag =&\left[ 1-2f_s(\omega) \right] \left[ \Sigma_s^\textrm{res,R}(\omega) - \Sigma_s^\textrm{res,R}(\omega)^\dagger \right]
\end{align}
Here, $W_s$ is given by 
\begin{equation} 
W_s(\omega)=\frac{1}{\sqrt{|\Delta_s^2-\omega^2|}} \left\{ \begin{array}{cr} 1 & |\omega| < \Delta_s \\ i \, 
\textrm{sgn}(\omega) & |\omega|>\Delta_s \end{array} \right. 
\end{equation}
and $f_s(\omega)=(e^{\omega/T_s}+1)^{-1}$ denotes the Fermi function in lead $s$. Of course, the same symmetries as for $V>0$ hold; the corresponding equations can be obtained by dropping the discrete Fourier indices in the equations above.

\section{Flow equations}
\label{sec:flow_eq}
The main idea of functional RG is to introduce a cut-off parameter $\Lambda$ into the free single-particle propagator $G^{\alpha \alpha^\prime}_\textrm{free} \to G_\textrm{free}^{\alpha \alpha^\prime,\Lambda}$ such that the single-particle irreducible vertex functions are known exactly at a particular $\Lambda_0$ and that $\Lambda=0$ corresponds to the original system. Now, a set of differential equations for the vertex functions is derived describing their ``flow'' from $\Lambda=\Lambda_0$ to $\Lambda=0$. These turn out to be an infinite set of coupled differential equations.\cite{Met12,Jak10c} The general flow equations and some further remarks can be found in App.\ \ref{sec:Details_fRG}. The initial conditions for the $n$-particle vertex functions are zero for $n>2$ if the Hamiltonian contains only two-particle interactions and the cut-off is chosen appropriately. The set of equations is truncated at the first (or second) order by setting the two- (or three-)particle vertex functions to their initial values throughout the entire flow. By this procedure, the entire first (or second) order of perturbation theory is captured and systematically enhanced in higher orders. A comprehensive presentation of the method in the context of the (normal-conducting) SIAM can be found in Refs.\ \onlinecite{Gez07}, \onlinecite{Jak10a} and \onlinecite{Jak10c}.

In this work, a hybridization flow parameter is introduced in analogy to Ref.\ \onlinecite{Jak10a}. It can be thought of as an additional artificial (metallic) reservoir that is coupled to the dot via a hybridization constant $\Lambda$, which assumes the role of the flow parameter flowing from $\infty$ to $0$. At the end of the flow, the additional reservoir is completely decoupled (as $\Lambda=0$) and the original system is obtained.

\subsection{Flow equations for $\mathbf{V>0}$}
For $V>0$, this implies the following additional hybridization self-energy:
\begin{align}
\Sigma_{\mathcal{H}}^{\textrm{R}}(\omega)^\Lambda_{n^\prime n}=- i \delta_{n^\prime n} \Lambda \left( \begin{array}{cc} 1 & 0 \\ 0 & 1 \end{array} \right)
\end{align}
\begin{align}
\Sigma_{\mathcal{H}}^{\textrm{K}}(\omega)^\Lambda_{n^\prime n}=\left[1-2f_\mathcal{H}(\omega_n)\right] \left[ \Sigma_{\mathcal{H}}^{\textrm{R}}(\omega)^\Lambda_{n^\prime n} - \Sigma_{\mathcal{H}}^{\textrm{A}}(\omega)^\Lambda_{n^\prime n}\right]
\end{align}
Here, $f_\mathcal{H}$ denotes the Fermi function and $T_\mathcal{H}=T_\textrm{L}=T_\textrm{R}$ is a reasonable choice as no temperature-gradients are investigated for $V>0$.

The one- and two-particle vertex functions (i.e. the self-energy due to the interaction $\Sigma$ and a renormalized two-particle interaction $\gamma$) will be the flowing quantities in the approximation schemes discussed here. They are parameterized in a way that they are not dependent on any continuous frequency arguments. However, they may carry discrete frequency indices accounting for the global periodicity of the problem. For both quantities the siFT is used whereas for propagators the diFT is used (in order to best exploit the respective advantages of the FTs). Two approximations were put to use: In each of them, the self-energy carries a single Fourier index $\Sigma(\Omega)^\Lambda_m \to \Sigma^\Lambda_m$---in time space, this corresponds to $\Sigma(t,t^\prime)\sim \Sigma(t) \delta(t-t^\prime)$ with periodic $\Sigma(t)$. In the simplest truncation scheme (\textSigma P1O), it is the only flowing quantity ($\gamma^\Lambda \to \pm U/2$). The name \textSigma P1O shall indicate that the approximation includes a periodic $\Sigma$ (\textSigma P) and is truncated such that the first order is captured completely (1O). In \textSigma P2O, also the two-particle vertex is renormalized ($\gamma^\Lambda \to \pm U_\Lambda/2$). In addition, a second order scheme that allows for a periodic $\gamma$ (\textgamma P2O) was derived, in which the two-particle vertex aquires an $m$-dependence ($\gamma^\Lambda \to \pm U^\Lambda_m/2$). As \textgamma P2O data is not shown here (due to the fact that convergence with respect to the numerical parameters could hardly be reached), this method is not discussed in detail.

The initial conditions of the flowing quantities are [note the remark after Eq.\ \eqref{eq:free_HF_propagator}; $\Omega,\Pi,X,\Delta$ denote bosonic frequencies associated with relative times while $m$ denotes the Fourier index associated with the centered time---for details see App.\ \ref{sec:FT}]:
\begin{equation} \label{eq:Anfbed_Sigma}
\Sigma^{\textrm{R}}(\Omega)^{\Lambda=\infty}_{m}=0
\end{equation}
\begin{align} \label{eq:Anfbed_vertex}
&\gamma_{q_1^\prime q_2^\prime q_1 q_2}^{\alpha_1^\prime \alpha_2^\prime \alpha_1 \alpha_2}(\Pi,X,\Delta)^{\Lambda=\infty}_{m}
\\=&\left\{ \begin{array}{cl} \delta_{m,0} \frac{1}{2} \bar{v}_{q_1^\prime q_2^\prime q_1 q_2} & \textrm{for } \alpha_1^\prime\! +\! \alpha_2^\prime\! +\! \alpha_1\! +\! \alpha_2 \! =\! \textrm{odd} \\ 0 & \textrm{else} \end{array} \right. \notag \end{align}
with:
\begin{equation}\bar{v}_{q_1^\prime q_2^\prime q_1 q_2}=\left\{ \begin{array}{cc} -U & q_1^\prime= q_1 \neq q_2= q_2^\prime \\ U & q_1^\prime= q_2 \neq q_1 =q_2^\prime \\ 0 & \textrm{else} \end{array} \right.\end{equation}

The derivation of \textSigma P1O is straight-forward and one finds:
\begin{align} \label{eq:Flow_eq_SigmaP1O}
\frac{\partial}{\partial \Lambda} \Sigma_{q^\prime q, m}^{\textrm{R}, \Lambda}
= i (-1)^{q-q^\prime} \frac{U}{2} \sum_n \int_{-V/2}^{V/2} \frac{d\omega}{2\pi}  S_{\bar{q} \bar{q}^\prime}^{\textrm{K}}(\omega)^\Lambda_{n,n-m}
\end{align}
Here, $S$ denotes the single-scale propagator which is calculated as $S^\Lambda=G^\Lambda (\partial \Sigma^\Lambda_\mathcal{H}/\partial \Lambda) G^\Lambda$, where the retarded component of the inverse full propagator is:
\begin{equation}
\left(G^{-1}\right)^{\textrm{R}}(\omega)^\Lambda_{n^\prime,n}=\left(G_\textrm{free}^{-1}-\Sigma^\textrm{res}-\Sigma_{\mathcal{H}}^\Lambda\right)^\textrm{R}(\omega)_{n^\prime,n} - \Sigma_{(n^\prime-n)}^{\textrm{R},\Lambda}
\end{equation}
In an analogous way, $\left(G^{-1}\right)^{\textrm{K},\Lambda}(\omega)_{n^\prime n}$ can be calculated (note that in all applied schemes, one finds $\Sigma^\textrm{K}=0$). Knowing these two quantities, $(G^{-1})^{\alpha^\prime \alpha,\Lambda}_{q^\prime q}(\omega)_{n^\prime n}$ can be inverted.

Deriving \textSigma P2O is more involved. The first step is neglecting the frequency dependence of the two-particle vertex by setting the external frequencies to zero in the general flow equation. In $V=0$ Matsubara functional RG, this step automatically yields a single real number $U_\Lambda$ describing the flowing two-particle vertex.\cite{Kar10} In contrast, this does not happen in Keldysh functional RG. In order to achieve this goal for finite $V$ and arbitrary physical parameters, the following procedure was applied: Those components that used to be zero at $\Lambda=\infty$ [see Eq.\ \eqref{eq:Anfbed_vertex}] are kept at zero. This leaves 32 components which partly can be linked to each other via symmetry relations down to four independent components, e.g.
\begin{equation}\gamma_{0101}^{0001,\Lambda},\gamma_{0101}^{0010,\Lambda},\gamma_{0101}^{1101,\Lambda},\gamma_{0101}^{1110,\Lambda}\end{equation}
The first infinitesimal step of the flow can be shown to yield:
\begin{align}
\notag \frac{\partial}{\partial \Lambda}\gamma_{0101}^{0001,\Lambda}=\frac{\partial}{\partial \Lambda} \gamma_{0101}^{1110,\Lambda},\; \frac{\partial}{\partial \Lambda}\gamma_{0101}^{0010,\Lambda}=\frac{\partial}{\partial \Lambda}\gamma_{0101}^{1101,\Lambda}\end{align}
This gives the motivation to apply the following mapping after each step of the flow in order to achieve the goal of a single real number describing the flow:
\begin{equation}\frac{\partial}{\partial \Lambda} U_\Lambda=-\frac{\partial}{\partial \Lambda} \textrm{Re}\left[\gamma_{0101}^{0001,\Lambda}+\gamma_{0101}^{0010,\Lambda} \right]\end{equation}
The set of flow equations is comprised of Eq.\ \eqref{eq:Flow_eq_SigmaP1O} with $U\to U_\Lambda$ and
\begin{widetext}
\begin{align}
\frac{\partial}{\partial \Lambda} U_\Lambda
= \textrm{Im} \left(\vphantom{\sum_{p^\prime}} \sum_{n n^\prime} \left(\frac{U_\Lambda}{2}\right)^2 \int_{-V/2}^{V/2} \frac{d\omega}{2 \pi}  \right.
&  \left\{ 2 \sum_{p p^\prime} (-1)^{p+p^\prime} \left[S^\textrm{A}_{p p^\prime} G^\textrm{K}_{\bar{p} \bar{p}^\prime} + S^\textrm{K}_{p p^\prime} G^\textrm{A}_{\bar{p} \bar{p}^\prime} \right](\omega)_{n,n^\prime} (-\omega)_{-n,-n^\prime} \right.
\\ \notag &\hphantom{xxx} +\left[\left(S^\textrm{R}_{0 0}+S^\textrm{A}_{0 0}\right) G^\textrm{K}_{1 1} + S^\textrm{K}_{0 0} \left(G^\textrm{R}_{1 1}+G^\textrm{A}_{1 1}\right) \right](\omega)_{n,n^\prime} (\omega)_{n^\prime,n} + S \leftrightarrow G 
\\ \notag &\hphantom{xxx} - \left. \left. \left[\left(S^\textrm{R}_{0 1}+S^\textrm{A}_{0 1}\right) G^\textrm{K}_{1 0} + S^\textrm{K}_{0 1} \left(G^\textrm{R}_{1 0}+G^\textrm{A}_{1 0}\right) \right](\omega)_{n,n^\prime} (\omega)_{n^\prime,n} + S \leftrightarrow G \vphantom{\sum_{p^\prime}} \right\} \right)
\end{align}
\end{widetext}
The notation $(f g)(\omega)_{n n^\prime}(\nu)_{m m^\prime}=f(\omega)_{n n^\prime} g(\nu)_{m m^\prime}$ was introduced here.
As indicated before, the \textSigma P1O set-up of differential equations corresponds to plain first order perturbation theory which is enhanced in a systematic way. This systematic way was derived from the general flow equations (which are exact) by the truncation and approximation considerations presented above. As the frequency dependence of the two-particle vertex was neglected in \textSigma P2O, it does not capture all terms of plain second order perturbation theory. Rather, it constitutes a more sophisticated resummation scheme than \textSigma P1O that is also complete to first order only.

\subsection{Flow equations for $\mathbf{V=0}$}
For $V=0$, the hybridization self-energy is:
\begin{align}
\Sigma_{\mathcal{H}}^{\Lambda,R}(\omega)=- i \Lambda \left( \begin{array}{cc} 1 & 0 \\ 0 & 1 \end{array} \right)
\end{align}
\begin{equation}
\label{eq:V_0_Sigma_hyb_K_neq} \Sigma_{\mathcal{H}}^{\textrm{K},\Lambda}(\omega)=- i 2 \left[1 - 2\left(\frac{\Gamma_\textrm{L}}{\Gamma}f_{\textrm{L}}(\omega)+\frac{\Gamma_\textrm{R}}{\Gamma}f_{\textrm{R}}(\omega)\right)\right] \Lambda \left( \begin{array}{cc} 1  & 0 \\ 0 & 1 \end{array} \right)
\end{equation}
The Keldysh component is more complicated than for $V>0$: In order to deal with the case of a finite temperature bias, two additional hybridization reservoirs are coupled to the dot (their coupling weighted by $\Gamma_s/\Gamma$), each of them being at the temperature of the corresponding lead. For $\Delta T=0$, this complicated structure collapses to the regular one.

As for $V>0$, the one- and two-particle vertex functions are the flowing quantities and their frequency dependence is neglected. In a static second order scheme (S2O), the two-particle vertex is parameterized by a single real number that flows additionally to the self-energy.

The initial conditions are:
\begin{equation} \label{eq:Anfbed_Sigma_V_0}
\Sigma^{\textrm{R},\Lambda=\infty}(\omega)=0
\end{equation}
\begin{align} \label{eq:Anfbed_vertex_V_0}
&\gamma_{q_1^\prime q_2^\prime q_1 q_2}^{\alpha_1^\prime \alpha_2^\prime \alpha_1 \alpha_2, \Lambda=\infty}(\omega_1^\prime,\omega_2^\prime,\omega_1,\omega_2)
\\=&\left\{ \begin{array}{cl} \delta(\omega_1^\prime+\omega_2^\prime-\omega_1-\omega_2)\frac{1}{2} \bar{v}_{q_1^\prime q_2^\prime q_1 q_2} & \textrm{for } \sum \alpha_{i}^{(\prime)} \! =\! \textrm{odd} \\ 0 & \textrm{else} \end{array} \right. \notag \end{align}

S2O is derived along the same lines as for $V>0$. Again, the same four independent components of $\gamma$ divide into two classes in the first infinitesimal step of the flow. Hence, $U_\Lambda$ is defined once more by averaging and taking the real part:
\begin{equation}\frac{\partial}{\partial \Lambda} U_\Lambda=-\frac{\partial}{\partial \Lambda} \textrm{Re}\left[\gamma_{0101}^{0001,\Lambda}+\gamma_{0101}^{0010,\Lambda} \right]\end{equation}
This procedure yields the flow equations:
\begin{align} \label{eq:Flow_eq_S1O}
\frac{\partial}{\partial \Lambda} \Sigma_{q^\prime  q}^{\textrm{R}, \Lambda}
= i (-1)^{q-q^\prime} \frac{U_\Lambda}{2} \int_{-\infty}^{\infty} \frac{d\nu}{2\pi}  S_{\bar{q} \bar{q}^\prime}^{\textrm{K}, \Lambda}(\nu)
\end{align}

\begin{align}
\label{eq:V0_flow_eq_S2O}
&\frac{\partial}{\partial \Lambda} U_\Lambda
= \textrm{Im} \left(\vphantom{\sum_{p^\prime}} \sum_{n n^\prime} \frac{U^2_\Lambda}{2} \int_{-\infty}^{\infty} \frac{d\omega}{2 \pi}  \right.
\\ \notag &\hphantom{xxx} \times \left\{ \sum_{p p^\prime} (-1)^{p+p^\prime} \left[S^\textrm{A}_{p p^\prime} G^\textrm{K}_{\bar{p} \bar{p}^\prime} + S^\textrm{K}_{p p^\prime} G^\textrm{A}_{\bar{p} \bar{p}^\prime} \right](\omega) (-\omega) \right.
\\ \notag &\hphantom{xxxxx} +\left[S^\textrm{A}_{0 0} G^\textrm{K}_{1 1} + S^\textrm{K}_{0 0} G^\textrm{R}_{1 1} \right](\omega) (\omega) + S \leftrightarrow G 
\\ \notag &\hphantom{xxxxx} - \left. \left. \left[\left(S^\textrm{R}_{0 1}+S^\textrm{A}_{0 1}\right) G^\textrm{K}_{1 0} + S^\textrm{K}_{0 1} \left(G^\textrm{R}_{1 0}+G^\textrm{A}_{1 0}\right) \right](\omega) (\omega) \vphantom{\sum_{p^\prime}} \right\} \right)
\end{align}

Once more, $\Sigma^\textrm{K}=0$. As above, the single-scale propagator is $S^\Lambda=G^\Lambda (\partial \Sigma^\Lambda_\mathcal{H}/\partial \Lambda) G^\Lambda$. As before, the frequency dependence of the two-particle vertex has been neglected and thus this set of equations corresponds to a sophisticated enhanced (in arbitrary high orders) form of first order perturbation theory. A static first order scheme could be obtained by setting $U_\Lambda=U$ instead of evolving it according to Eq.\ \eqref{eq:V0_flow_eq_S2O}.

\section{Formula for the current and perturbation theory}
\label{sec:Formula_current}

Here, we are interested in the current as the observable. Setting the electronic charge equal to $-1$, one finds for the current going into reservoir $s$ that $J_s(t)=-i\langle[H(t),N_s(t)]\rangle$. 

\subsection{Current and SCHF for $\mathbf{V>0}$}
Calculating the commutator on the right-hand-side and identifying Keldysh Green functions in the resulting terms, one finds a periodic time-dependence of the current $J_{s}(t)=\sum_\nu e^{-i\nu V t} \hat{J}_{s,\nu}$ for $V>0$.
The following formula can be derived for $\hat{J}_{s,\nu}=(\hat{j}_{s,\nu}+\hat{j}_{s,-\nu}^\ast)/2$:
\begin{align}
\notag \hat{j}_{s,\nu} = - \sum_{n n^\prime} \int_{-V/2}^{V/2}  \frac{d \omega}{2 \pi}
& \textrm{Tr}\left( \sigma_z \left[ \Sigma_{s}^{\textrm{res,K}}(\omega)_{n+\nu,n^\prime} G^\textrm{A}(\omega)_{n^\prime,n} \right. \right.
\\ & \hphantom{xxx}\left. \left. + \Sigma_{s}^{\textrm{res,R}}(\omega)_{n+\nu,n^\prime} G^\textrm{K}(\omega)_{n^\prime,n} \right]\right)
\end{align}
We will focus on the dc-current $J_{s,0}$. Of course, current conservation $J_{\textrm{L},0}=-J_{\textrm{R},0}$ must hold in exact calculations---note that this conservation cannot be proven to be fulfilled for the truncated functional RG scheme proposed above in combination with this current formula. However, the results we present conserve the current $I_\nu=J_{\textrm{L},\nu}=-J_{\textrm{R},\nu}$.

Some restricted self-consistent Hartree-Fock (SCHF) results will be shown. They were calculated along the lines of Ref.\ \onlinecite{Del08}. The following equation is iterated until $\Sigma^\textrm{R}_{q^\prime q,m}$ is converged numerically:
\begin{equation}
\Sigma^\textrm{R}_{q^\prime q,m}=i (-1)^{q+q^\prime} \frac{U}{2} \sum_n \int_{-V/2}^{V/2}\frac{\textrm{d}\omega}{2\pi} G^\textrm{K}_{\bar{q}\bar{q}^\prime}(\omega)_{n,n-m}
\end{equation}

\subsection{Current for $\mathbf{V=0}$}
For $V=0$, the same derivation as for $V>0$ yields a time-independent current:
\begin{align} \label{eq:V_0_formula_current}
J_{s}=\textrm{Re}\left\{-\int_{-\infty}^{\infty} \frac{d \omega}{2 \pi} \textrm{Tr}\left( \sigma_z \left[ \vphantom{\Sigma^K} \right. \right. \right. & \Sigma_{s}^{\textrm{res,K}}(\omega) G^{\textrm{A}}(\omega) +
\\ \notag & \left.\left.\left. \Sigma_{s}^{\textrm{res,R}}(\omega) G^{\textrm{K}}(\omega)\right]\right) \vphantom{\int_\infty^\infty}\right\}
\end{align}
The same problem with current conservation as for $V>0$ occurs; results shown here do not violate current conservation.

\section{Numerical results}
\label{sec:num_results}

Although our method does not require these limitations, we will restrict the discussion to $\Gamma_\textrm{L}=\Gamma_\textrm{R}=\Gamma/2$, $\Delta_\textrm{L}=\Delta_\textrm{R}=\Delta$, $B=0$ and $\phi_\textrm{L}=-\phi_\textrm{R}=\phi/2$. As indicated above, it is advantageous to exploit the $B=0$ symmetries.\footnote{For $V=0$, the code becomes even numerically unstable unless the $B=0$ symmetries are imposed explicitly.}

\subsection{Results for $\mathbf{V>0}$}
\label{sec:Results_Vn0}

For numerical $V>0$ calculations , the continuous frequency $\omega \in [-V/2,V/2)$ must be discretized. A numerical parameter $\Omega_\textrm{len}$ is introduced ($\Omega=-\Omega_\textrm{len},..,\Omega_\textrm{len}$): $\omega_\Omega=V \Omega /(2\Omega_\textrm{len}+1)$. For each single-particle quantity (treated by diFT) and each $\omega_\Omega$, two matrices ($^\textrm{R}$ and $^\textrm{K}$) are stored containing the indices $q,q^\prime$ and $n,n^\prime$. Also, a cut-off $m_\textrm{len}$ is introduced for the $n,n^\prime=-m_\textrm{len},..,m_\textrm{len}$. This is also the range for the siFT index $m=-m_\textrm{len},..,m_\textrm{len}$. Of course, convergence with respect to $m_\textrm{len}$ and $\Omega_\textrm{len}$ must be checked---which will turn out to be a major issue in some cases. Note that every second element of the matrices induced by the superindices $M^{(\prime)}=q^{(\prime)}(2 m_\textrm{len}+1)+(n^{(\prime)}+m_\textrm{len})$ can be shown to vanish (in the approximations discussed); this is exploited in numerical calculations.

We will focus on two sets of parameters (both at rather small $U/\Gamma$ and at $T=T_\textrm{L/R}=0$ as well as $\phi=0$)---one of which exhibits a large influence of MAR and one of which does not.

\begin{figure}
\includegraphics[width=0.23\textwidth]{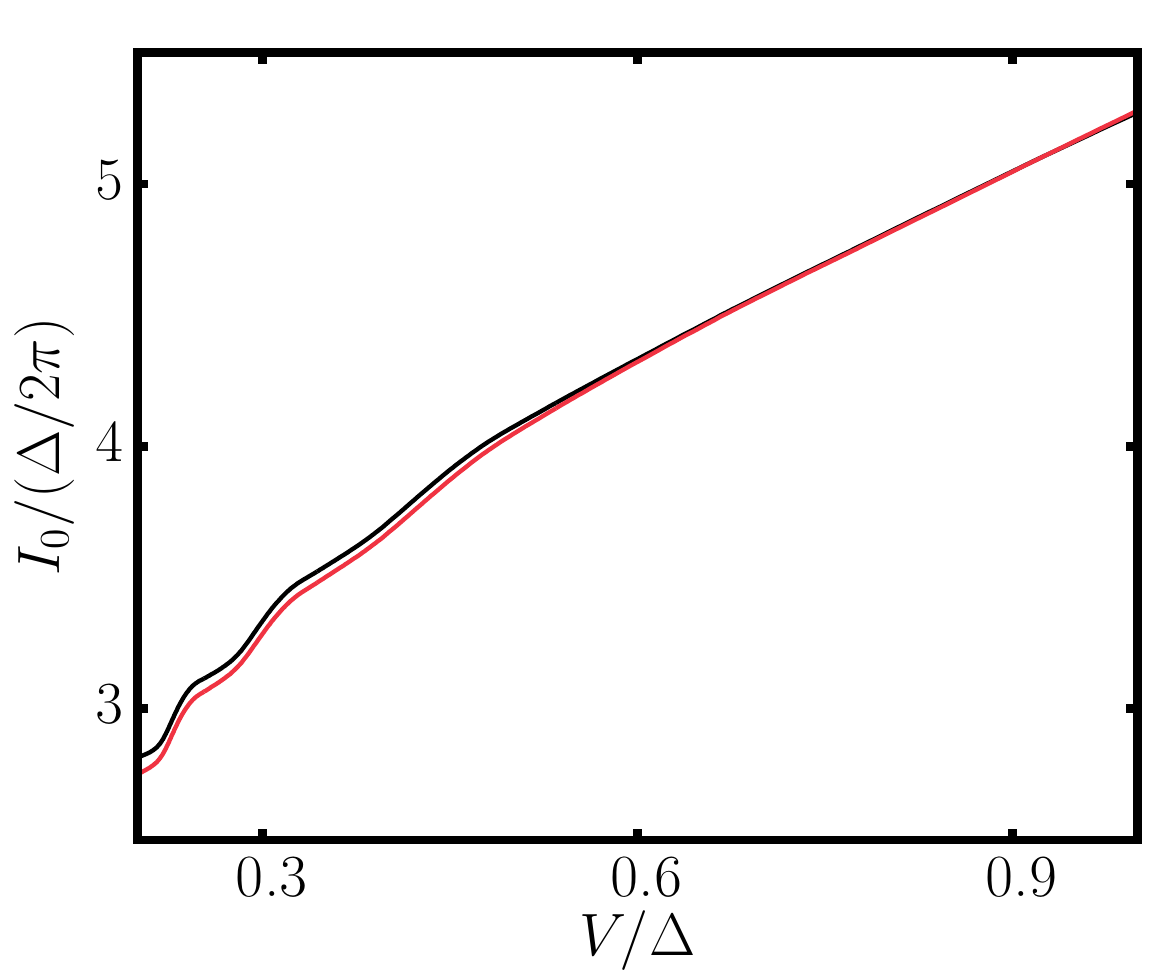}
\includegraphics[width=0.23\textwidth]{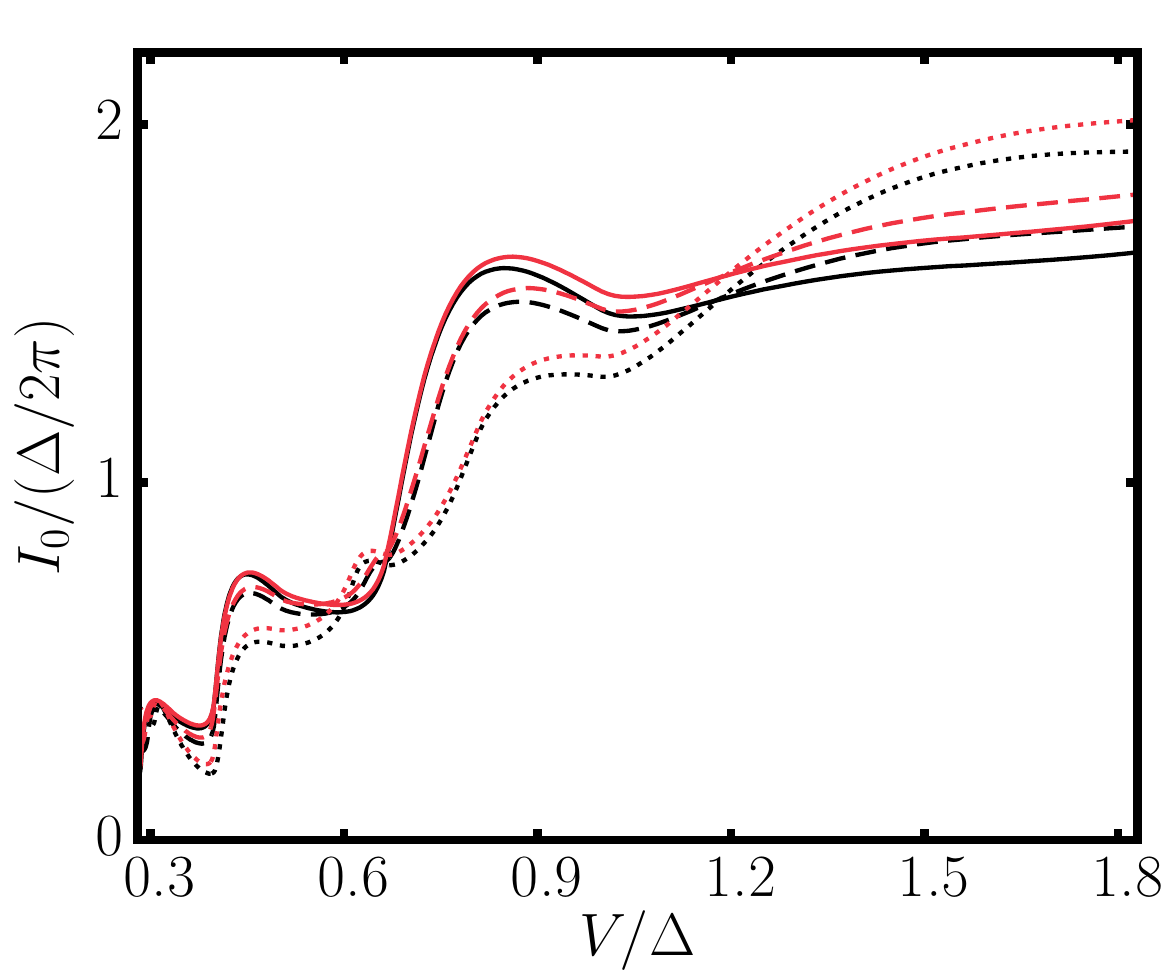}
\caption{\label{fig:00} (Color online) These plots show numerical data for the dc-component of the current for the two parameter sets investigated in Sect.\ \ref{sec:Results_Vn0}. Black lines are $U=0$ results, whereas red (or grey) lines are \textSigma P1O results.
The left plot has $U/\Gamma=0.25$, $\Delta/\Gamma=0.25$, $V_\textrm{g}/\Gamma=0$ and $\Omega_\textrm{len}=96$, $m_\textrm{len}=90$.
The right plot has $U/\Gamma=0.25$, $\Delta/\Gamma=1$ and $\Omega_\textrm{len}=96$, $m_\textrm{len}=64$. $V_\textrm{g}/\Gamma=0.0$ is solid, $V_\textrm{g}/\Gamma=0.2$ is dashed and $V_\textrm{g}/\Gamma=0.4$ is dotted.}
\end{figure}

\begin{figure}
\includegraphics[width=0.45\textwidth]{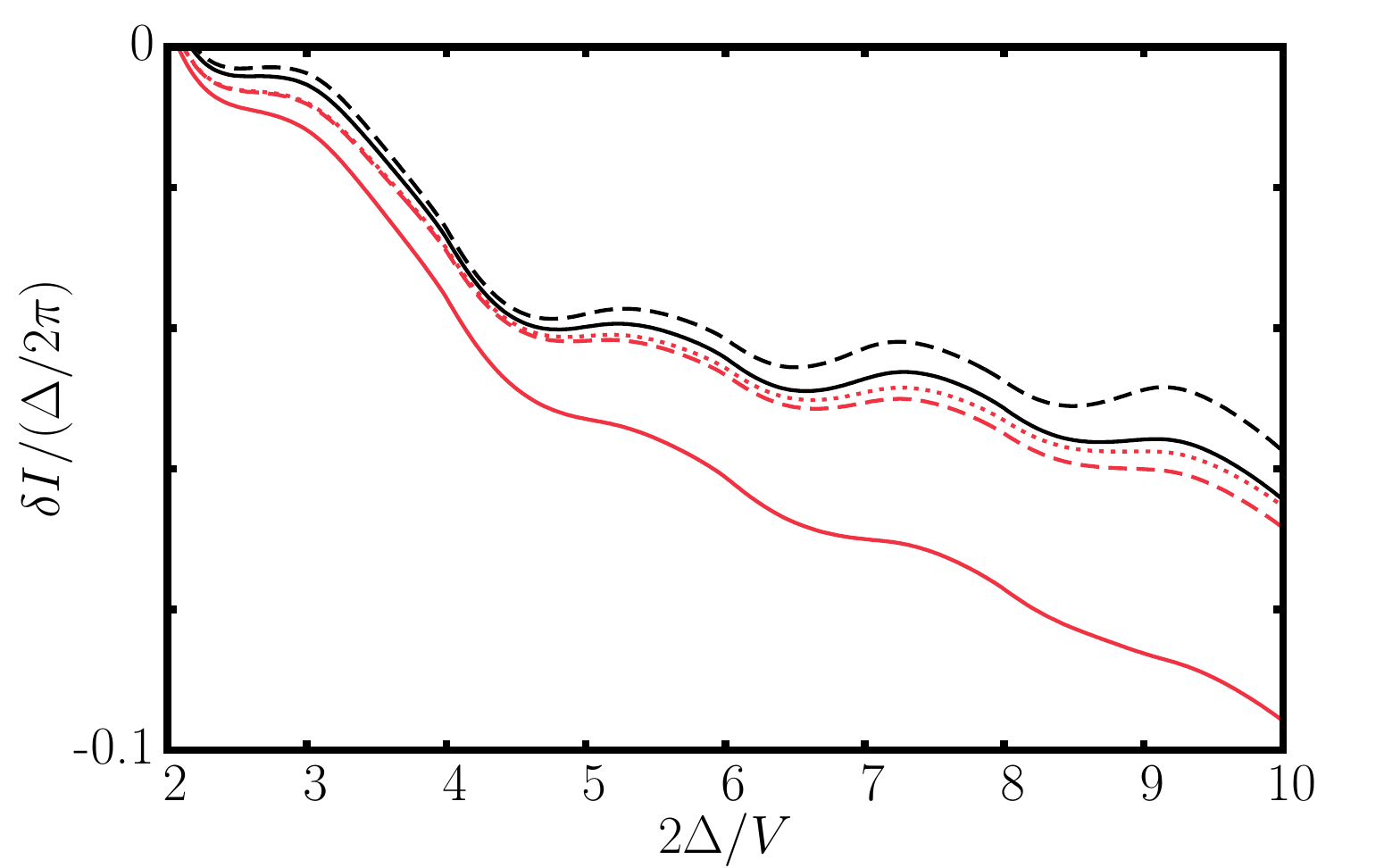}
\caption{\label{fig:01} (Color online) This plot shows numerical data for $U/\Gamma=0.25$, $\Delta/\Gamma=0.25$ and $V_\textrm{g}/\Gamma=0$ (and $\Omega_\textrm{len}=96$). The red (or grey) lines show SCHF data for $m_\textrm{len}=24, 90, 150$ from bottom to top. This illustrates that convergence is (only) reached up to $2 \Delta/V \approx 5$ for $m_\textrm{len}=90$. The black solid line is \textSigma P1O and the black dashed line is \textSigma P2O data at this $m_\textrm{len}=90$.}
\end{figure}

The first choice of parameters is $U/\Gamma=0.25$, $\Delta/\Gamma=0.25$ and $V_\textrm{g}/\Gamma=0$. Some numerical results for the dc-current $I_0$ are shown in the left part of Fig.\ \ref{fig:00}. The difference between the interacting and the non-interacting dc-current $\delta I=I_0(U>0)-I_0(U=0)$ is shown in Fig.\ \ref{fig:01}. The interaction moderately suppresses the current. In the present case of broad dot levels $\Gamma>\Delta$, MAR play a subdominant role and the interaction correction to the dc-current only shows weak MAR features. Reaching convergence with respect to $m_\textrm{len}$ is a problematic issue for this set of parameters. This problem is illustrated by the red (or grey) SCHF data. For $2 \Delta/V < 5$, convergence was achieved (see the $m_\textrm{len}=150$ data in comparison to the $m_\textrm{len}=90$ data). For $2\Delta/V > 5$, the $m_\textrm{len}=150$ data is still a little above the $m_\textrm{len}=90$ data. From the evolution of the discrepancy between the curves for increasing $m_\textrm{len}$, we estimate that convergence can be expected to have been reached up to $2\Delta/V \approx 7$ for $m_\textrm{len}=150$. This analysis implies that the best functional RG data we can show (at $m_\textrm{len} = 90$) cannot be expected to be fully converged for $2\Delta/V > 5$. Nevertheless, the main point can still be made because further increasing $m_\textrm{len}$ can only be expected to yield even higher curves: Our investigation yields a moderate suppression of the current due to the interaction. Note that in our work the \textSigma P2O curve is above the \textSigma P1O curve (which means that the current is even less suppressed).
Comparing to the SCHF result of Ref.\ \onlinecite{Del08} (which appears to be close to our $m_\textrm{len}=24$ SCHF data) we speculate that convergence with respect to $m_\textrm{len}$ has not been reached there. As this SCHF result then enters the calculation of the second order perturbation theory of Ref.\ \onlinecite{Del08} this second order result becomes questionable as well.

\begin{figure}
\includegraphics[width=0.45\textwidth]{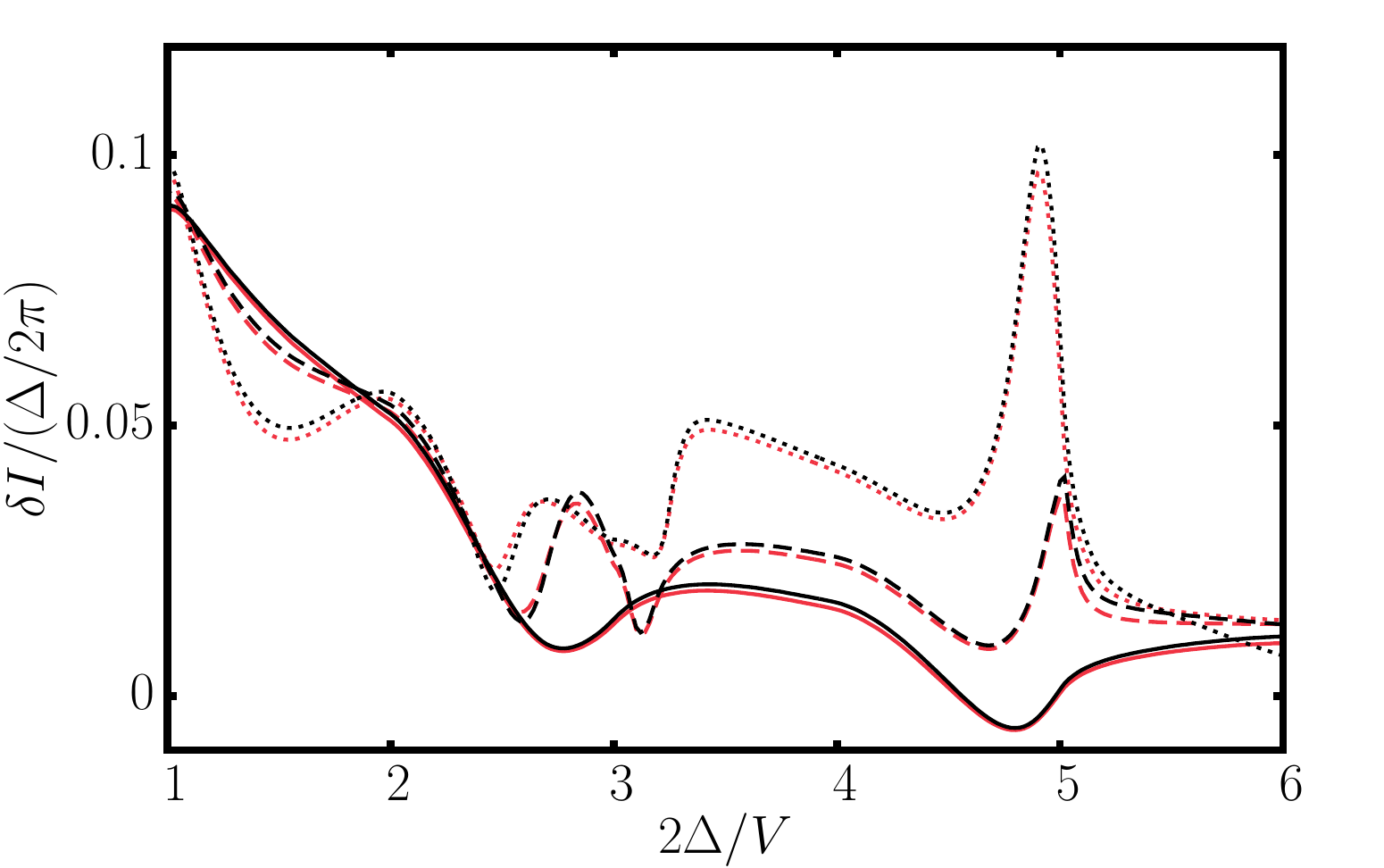}
\caption{\label{fig:02} (Color online) This plot shows numerical data for $U/\Gamma=0.25$, $\Delta/\Gamma=1$ and $V_\textrm{g}/\Gamma=0.0, 0.2, 0.4$ (and $m_\textrm{len}=64$, $\Omega_\textrm{len}=96$). Black lines show \textSigma P1O functional RG data, whereas the red (or grey) lines show SCHF data. $V_\textrm{g}/\Gamma=0.0$ is solid, $V_\textrm{g}/\Gamma=0.2$ is dashed and $V_\textrm{g}/\Gamma=0.4$ is dotted.}
\end{figure}

Now, a second set of parameters with larger superconducting gap $\Delta/\Gamma$ shall be discussed where MAR are important and convergence with respect to $m_\textrm{len}$ is not a problematic issue: $U/\Gamma=0.25$, $\Delta/\Gamma=1$, $V_\textrm{g}/\Gamma=0.0, 0.2, 0.4$. Numerically converged \textSigma P1O and SCHF results for $\delta I$ can be seen in Fig.\ \ref{fig:02}---see the right part of Fig.\ \ref{fig:00} for the $I_0$ curves. Two points can be made. First, \textSigma P1O and SCHF agree very well quantitatively---as both can be understood as enhanced first order schemes and the interaction is small enough such that SCHF does not show a spurious spin-symmetry breaking, this is plausible. Second, distinct interaction effects (at least for $V_\textrm{g}>0$) at the odd MAR points are observed---as explained in Sect.\ \ref{sec:Introduction}, this can be understood from a physical perspective. Thus, it is of particular interest whether \textSigma P2O functional RG can contribute to a better understanding---we investigate this question for the case with the least interaction effects at \textSigma P1O, namely $V_\textrm{g}/\Gamma=0$.
\begin{figure}
\includegraphics[width=0.45\textwidth]{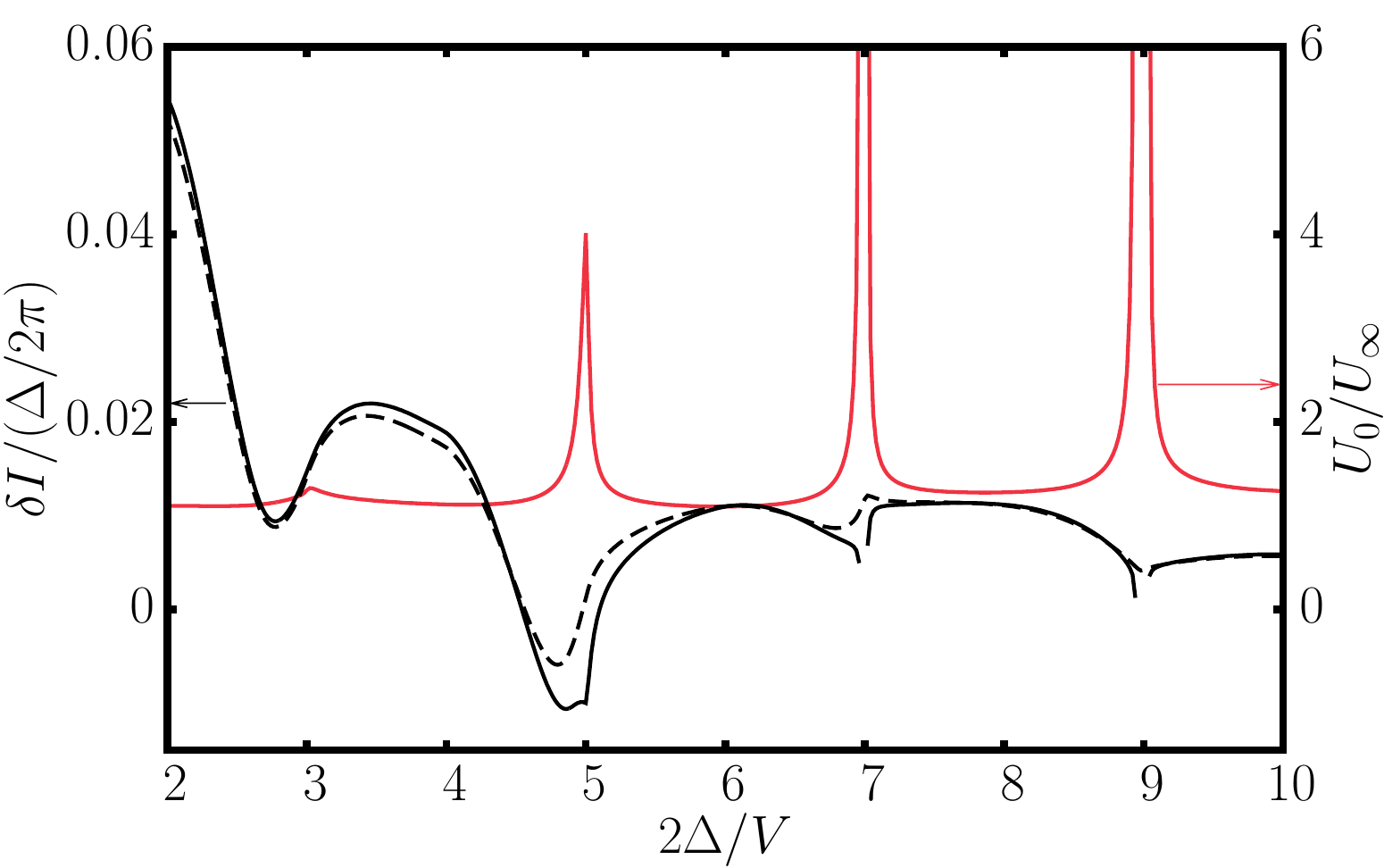}
\caption{\label{fig:03} (Color online) This plot shows numerical data for $U/\Gamma=0.25$, $\Delta/\Gamma=1$ and $V_\textrm{g}/\Gamma=0.0$ (and $m_\textrm{len}=64$, $\Omega_\textrm{len}=96$). The dashed line is \textSigma P1O and the solid one is \textSigma P2O. The renormalized two-particle vertex at the end of the \textSigma P2O flow $U_0/U_\infty$ is shown in red (or grey).}
\end{figure}
The numerical results are shown in Fig.\ \ref{fig:03}. \textSigma P2O shows a break-down of the method at the odd MAR points---this can be seen in the renormalized interaction divided by the bare one: $U_0/U_\infty$. This quantity serves as an indicator for the strength of interactions effects. It should take moderate values as it is the case for small $\Delta/\Gamma$ (e.g. in Fig.\ \ref{fig:01}). The existence of such an indicator (and thus an internal consistency check) is a very important feature of the method which in itself constitutes an advance compared to earlier methods applied to the problem. Such drastically increased values as seen in the vicinity of the odd MAR points in Fig.\ \ref{fig:03} show that the interaction effects are very strong and cannot be captured by a low order truncation functional RG scheme. In particular, the \textSigma P2O resummation scheme is not sufficient to prevent the growing of the renormalized two-particle interaction at the odd MAR points. Note that the RG flow does not even come to an end right around $2\Delta/V\approx 7,9$. A remaining question is whether a more sophisticated parametrization and truncation functional RG scheme would be able to avoid this break-down of the method. The calculations carried out with \textgamma P2O showed that (although convergence with respect to $m_\textrm{len}$ had not been reached yet) also this procedure is not sufficient to overcome the problems at the odd MAR points. Reviewing our results (and taking into account the internal consistency check) for \textSigma P1O, \textSigma P2O (and \textgamma P2O; not shown) as well as those obtained by SCHF and second order perturbation theory and combining them with the physical picture of a newly opening MAR channel, we argue that all methods that use an approach of perturbative character in $U$ are prone to problems at the odd MAR points for $\Gamma \lesssim \Delta$.

\subsection{Results for $\mathbf{V=0}$}
For $V=0$ and within the S2O approximation, the frequency integrations on the right-hand-sides of the flow equations can be carried out by continuous integration routines, i.e. no a priori discretization of the frequency axis is necessary. However, $\eta$ must be kept finite and numerical convergence for $\eta \to 0$ must be checked. In particular, in the vicinity of the broadened ABS (the zeros of the real part of the denominator of the retarded Green function), the numerical integration must be performed very carefully.
All numerical results shown here have $V_\textrm{g}=0$.

For $\Delta T=0$, the physics that is to be expected for finite interactions is well-known.\cite{Gla89,Sia04,Cho04,Kar08,Men09} In the case of $T=0$, a first order quantum phase transition leads to a sharp change of sign and amplitude of the current as a function of the complex phase difference of the superconductors $\phi=\phi_\textrm{L}-\phi_\textrm{R}$. The phase of $\phi$ smaller than the position of the phase transition $\phi_\textrm{c}$ is called singlet (or $0$) phase. The other phase at $\phi$ greater than $\phi_\textrm{c}$ is called doublet (or $\pi$) phase. Although the clear distinction breaks down for $T>0$ and $\Delta T \neq 0$ we still use the terms singlet and doublet phase to refer to the respective regions---a precise definition of the boundary is not required for our assertions. The $T=0$ equilibrium problem was studied with the Hartree-Fock method\cite{Mar11,Roz99,Yos00} which has the short-coming that the phase transition sets in due to an unphysical breaking of spin-symmetry. In contrast, Matsubara functional RG predicts a phase transition without breaking of spin-symmetry and was successfully used to investigate the $\Delta T=0$ equilibrium situation.\cite{Kar08,Eic09,Kar10,Kar11} The Keldysh functional RG proposed here is not equivalent to that approach (for $\Delta T=0$) but can be made so (at least in first order truncation) by replacing the hybridization cut-off with the analogue to the sharp imaginary frequency cut-off employed in the Matsubara functional RG (for details regarding this replacement see Ref.\ \onlinecite{Jak10c}).

\begin{figure}
\includegraphics[width=0.45\textwidth]{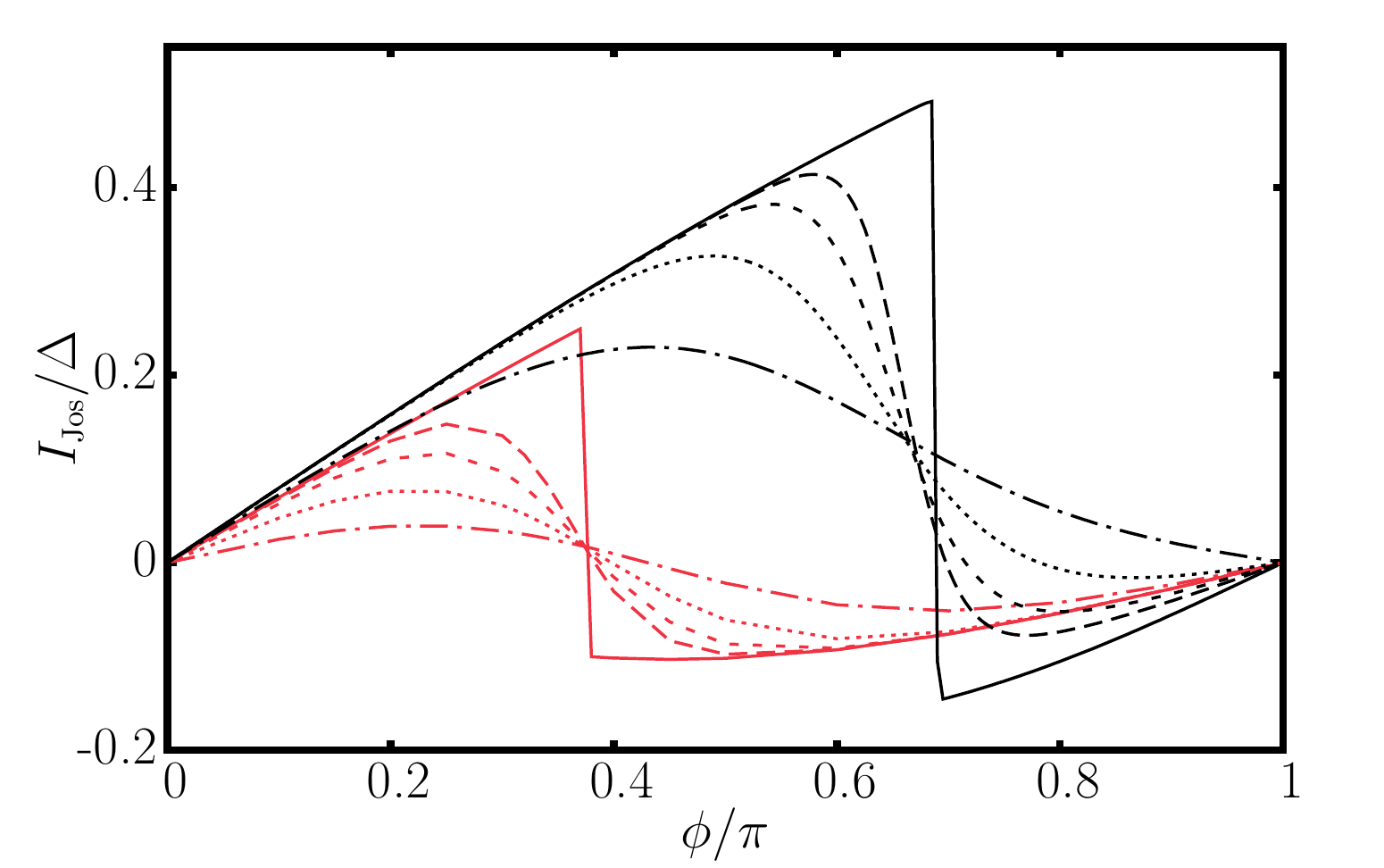}
\caption{\label{fig:04} (Color online) This plot shows numerical data for $\Delta/\Gamma=0.37$, $U/\Gamma=5.2$ and $\Delta T=0$ (and $\eta=10^{-4}$). NRG data is red (or grey) and S2O data is black. The solid curves with the sharp feature are at $T=0$, while the others have increasing temperatures $T/\Delta=0.02, 0.03, 0.05, 0.1$ up to the dashed-dotted line.}
\end{figure}

The S2O Keldysh functional RG method proposed above yields the $\Delta T=0$ physics very well qualitatively. This is illustrated in Fig.\ \ref{fig:04} where S2O data is compared to NRG data taken from Ref.\ \onlinecite{Kar08} which is expected to be very accurate. Note that in general the parameters must be fine-tuned for the phase transition to occur for varying $\phi$. The available NRG curves were calculated at a rather large $U/\Gamma=5.2$. The rest of the parameters are $\Delta/\Gamma=0.37$  and various $T/\Delta \in [0,0.1]$.  For this particular choice of parameters, S2O does not reproduce $\phi_\textrm{c}$ at $T=0$ very accurately. Typically, $\phi_\textrm{c}$ is strongly parameter-dependent (which is plausible if a fine-tuning as mentioned above is necessary). Thus, it is not surprising that the position is not reproduced exactly by an approximative method, especially at such large $U/\Gamma$. Consequently, we consider the ability to reproduce the position $\phi_\textrm{c}$ not to be a decisive criterion to judge the accuracy of a given approximate approach. The qualitative features of the $T>0$ curves are reproduced well by S2O. Also, the S2O $T>0$ curves have a common intersection point (as do the NRG curves).
In the doublet phase at $T=0$, the functional RG predicts a different current amplitude than NRG which can be traced back to an artefact of the functional RG method: The off-diagonal (i.e. ``superconducting'') self-energy component $\Sigma_{01}^\textrm{R}$ gets pinned to the value $\Gamma \textrm{cos}(\phi/2)$. This is an effect known from (equilibrium) Matsubara functional RG.\cite{Kar08} It forces the current on a universal curve dependent only on $\Gamma$ and $\Delta$, but not on $U$ or $V_\textrm{g}$. This is in contrast to the NRG curves which are (weakly) $U$ and $V_\textrm{g}$ dependent. In spite of these short-comings, S2O Keldysh functional RG captures the essential physics quite well in the equilibrium case. For smaller $U/\Gamma$, the method should be trusted even more.

\begin{figure}
\includegraphics[width=0.45\textwidth]{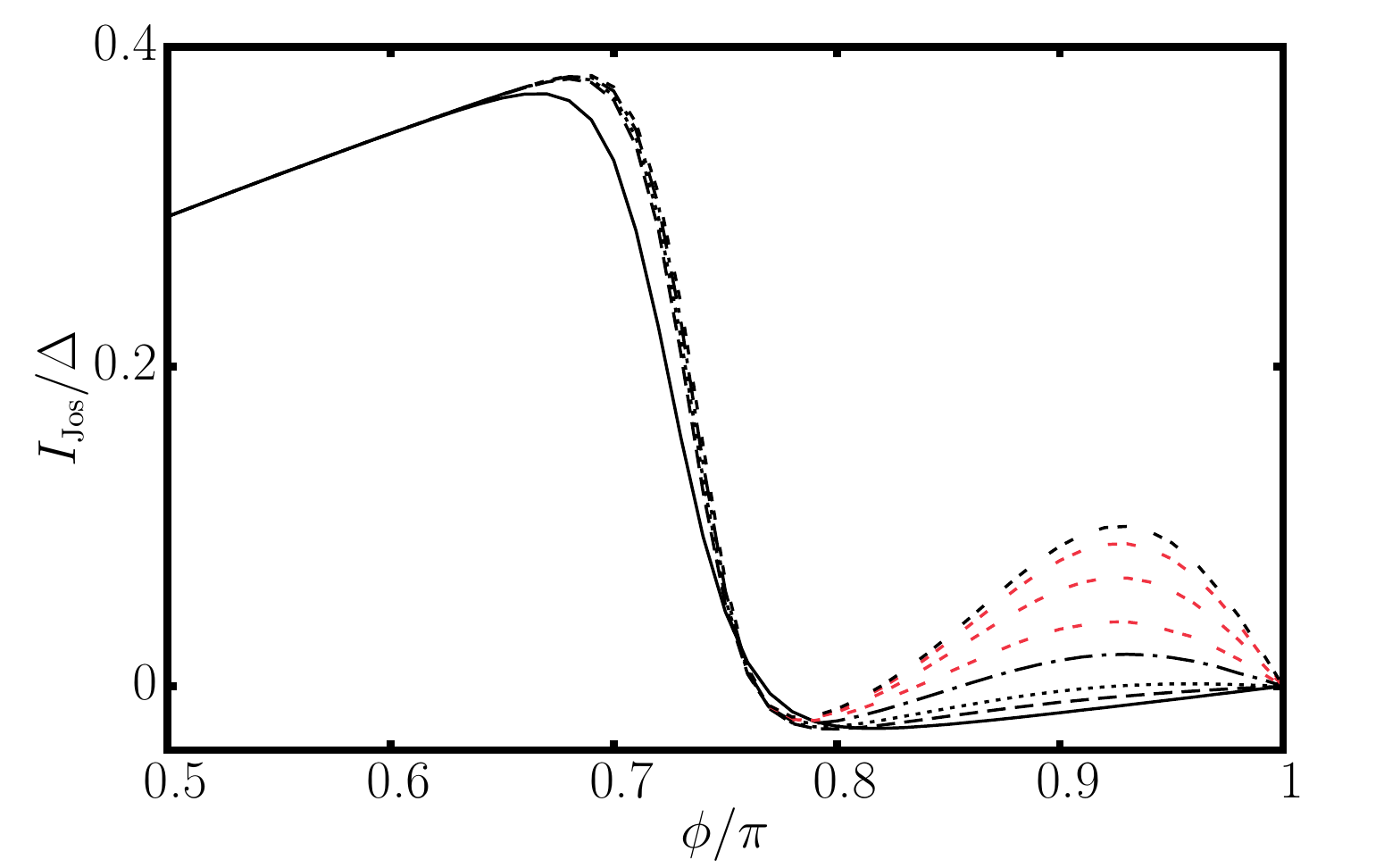}
\caption{\label{fig:05} (Color online) This plot shows current data for $\Delta/\Gamma=1$, $U/\Gamma=2$ and $1000 T_\textrm{L}/\Gamma=10$ obtained with S2O. $1000 T_\textrm{R}/\Gamma$ takes values of $10, 5, 4.5, 4, 3.5$ going from the solid to the double-dashed line.
Convergence with respect to $\eta$ is achieved with $\eta=10^{-4}$ for $\phi < 0.78 \pi$. For $\Delta T=0$ this is also sufficient for $\phi \geq 0.78 \pi$, while the other curves have $\eta=3.33\cdot 10^{-6}$. Convergence is increasingly difficult to achieve; this is illustrated by the $\eta=1.00\cdot 10^{-5}, 3.33\cdot 10^{-5},1.00\cdot 10^{-4}$ curves shown for $T_\textrm{R}/T_\textrm{L} = 3.5/10$ in red (or grey).}
\end{figure}

After this benchmarking of the method, we proceed to non-equilibrium induced by $\Delta T \neq 0$---the true purpose that Keldysh functional RG has been developed for. Also, to be on the safe side we consider a different set of parameters with smaller $U/\Gamma=2$ (and $\Delta/\Gamma=1$). A surprising increase of the current in the doublet phase can be observed if one starts at the $T/\Gamma=0.01$ equilibrium case and tunes $T_\textrm{R}/T_\textrm{L}$ to a non-equilibrium value of $\approx 1/2$ keeping $T_\textrm{L}$ fixed. This can be seen in Fig.\ \ref{fig:05}. For an increasing temperature gradient, the current in the doublet phase increases. Note that it does not matter whether $T_\textrm{L}$ or $T_\textrm{R}$ is kept fixed: it was found that $I_\textrm{Jos}(\phi;T_\textrm{L},T_\textrm{R})=I_\textrm{Jos}(\phi;T_\textrm{R},T_\textrm{L})$ holds numerically for the investigated parameters. The larger the effect, the more difficult it is to reach convergence with respect to $\eta$. This is shown for $T_\textrm{R}/T_\textrm{L}=3.5/10$. The smallest $\eta$ that could be reached is $3.33\cdot 10^{-6}$.
For the $T_\textrm{R}/T_\textrm{L}$ shown here, $U_0/U_\infty$ takes moderate values $\leq 1.2$. Further reducing $T_\textrm{R}/T_\textrm{L}$ leads to a significant increase of $U_0/U_\infty$. Again, this indicates the emergence of strong correlation effects and for more extreme temperature gradients the S2O resummation scheme is not sufficient to avoid the break-down.
This might raise doubts whether the observed effect is an artefact of the method. We emphasize that the effect occurs significantly at acceptable $U_0/U_\infty$. Furthermore, a current increase can even be observed in the simpler truncation scheme without renormalization of the two-particle vertex.

Several checks were performed to test whether the effect persists. A very important check is whether $J_\textrm{L}=-J_\textrm{R}$ as the current formula does not guarantee this symmetry in our truncation (as discussed in Sect.\ \ref{sec:Formula_current})---the equality does indeed hold numerically. It was checked whether the effect relies on the (symmetric) choices of the parameters. It was found that it does not vanish (at least not immediately) if one goes away from the symmetric choices of $V_\textrm{g}/\Gamma=0$, $\Gamma_\textrm{L}/\Gamma_\textrm{R}=1$ or $\Delta_\textrm{L}/\Delta_\textrm{R}=1$. Furthermore, the effect was also observed if one starts from the set of parameters from above (with $U/\Gamma=5.2$) and tunes $T_\textrm{R}/T_\textrm{L}$ to $\approx 1/2$.
Also, we checked whether the effect is due to numerical inaccuracies. Possible causes are too high upper error bounds for the numerical integration routines, too low integration limits (when performing integrals that formally go along the entire real axis) or an optimization procedure exploiting the knowledge of the numerically determined position of the ABS. None of these was found to be the reason for the effect.

\begin{figure}
\includegraphics[width=0.45\textwidth]{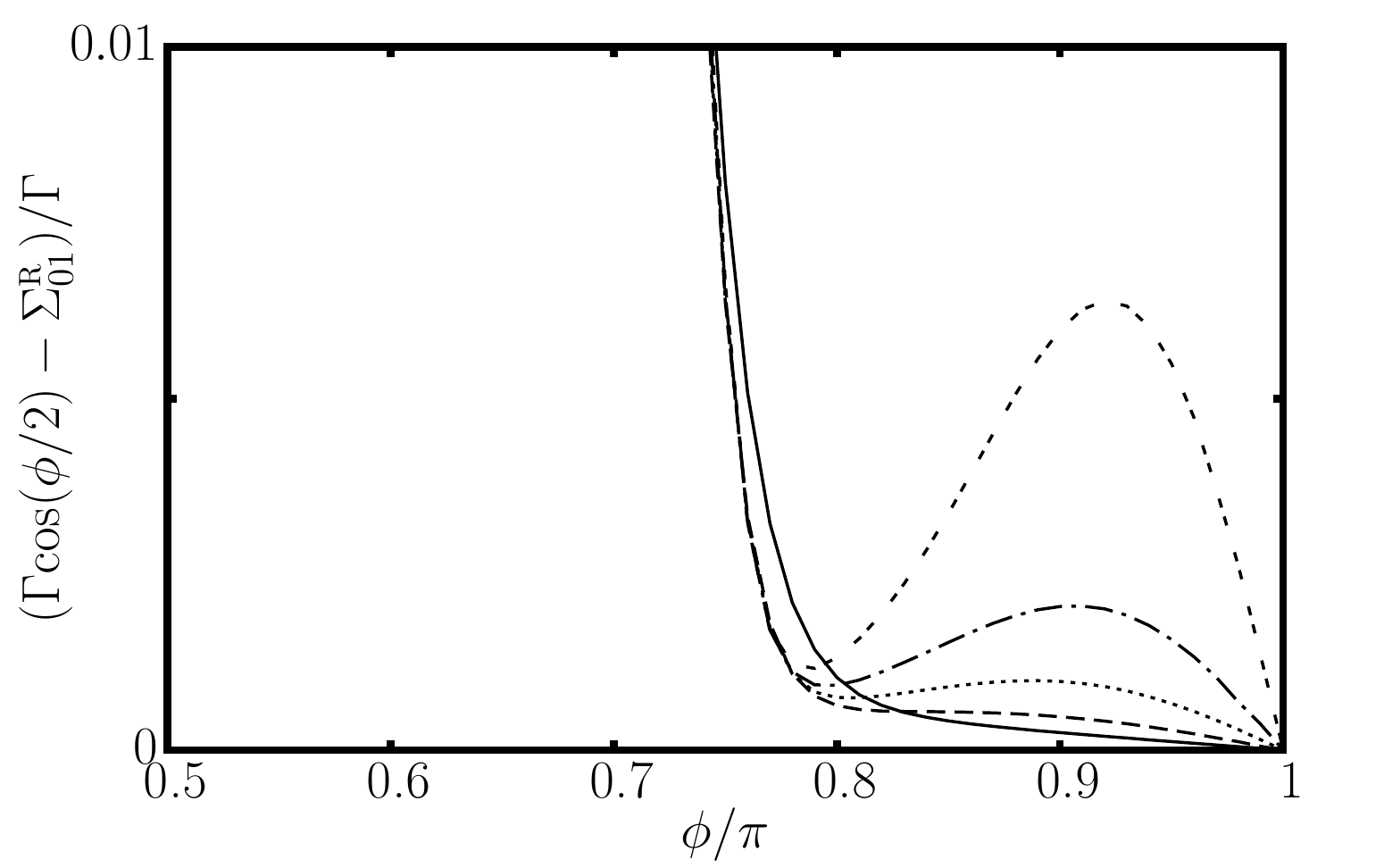}
\caption{\label{fig:06} This plot shows $(\Gamma \textrm{cos}(\phi/2)-\Sigma_{01}^\textrm{R})/\Gamma$ for the same parameters as in Fig.\ \ref{fig:05}.}
\end{figure}

The effect was traced back to the behavior of $\Sigma_{01}^\textrm{R}(\phi)$ in the doublet phase. Remember that $\Sigma_{01}^\textrm{R}(\phi)$ takes the value $\Gamma \textrm{cos}(\phi/2)$ at $T=0$. At $T_\textrm{L}=T_\textrm{R} > 0$, $\Sigma_{01}^\textrm{R}(\phi)$ starts to deviate from $\Gamma \textrm{cos}(\phi/2)$. This deviation increases significantly for $\Delta T \neq 0$---see Fig.\ \ref{fig:06}. This is the decisive ingredient in the current formula to produce the current increase effect. We observed semi-analytically that even the equilibrium Matsubara current formula reacts correspondingly to a deviation in $\Sigma_{01}^\textrm{R}(\phi)$.

\section{Conclusion}
\label{sec:conclusion}
We investigated the Josephson quantum dot in non-equilibrium. This non-equilibrium was either induced by a finite bias voltage or a temperature bias. The two cases had to be distinguished as a finite bias voltage implies a time-dependent Hamiltonian. In both cases, a Keldysh functional RG approach was employed; in its most sophisticated form, a static (i.e. not frequency-dependent) flow of the two-particle vertex was included. For the case of finite bias voltages, also the self-consistent Hartree-Fock method was used. We investigated the (dc-/Josephson) current as the observable.

Two sets of parameters were investigated for finite bias voltages. For a set of parameters where multiple Andreev reflections do not play an important role (small $\Delta/\Gamma$), we observe that numerical convergence is increasingly difficult to reach for increasing $\Delta/V$. However, this is still feasible within the self-consistent Hartree-Fock method. From the accuracy achieved overall, we can judge that the interaction induced supression of the current is much less significant than suggested by earlier works.\cite{Del08} In the second set of parameters (at larger $\Delta/\Gamma$), multiple Andreev reflections have a large influence on the current at the so called odd MAR points. Correlation effects at these points were found to be very strong. As a consequence, the flow scheme allowing for a renormalization of the self-energy as well as the two-particle vertex was found to become uncontrolled in the vicinity of these points. We argued that all perturbative methods (in the interaction) will be prone to problems at larger $\Delta/\Gamma$ due to multiple Andreev reflections. Consequently, the data they produce cannot be trusted close to the odd MAR points. This provides reason to pursue other than weak coupling approaches.

For vanishing bias voltage, we used an equilibrium set of parameters for which numerical RG data is available to benchmark the method. Then, we proceeded to a set of parameters with a smaller interaction parameter and induced non-equilibrium by tuning the ratio of right lead temperature to left lead temperature to about one half. A current increase effect was observed in the regime that used to be the doublet phase at vanishing lead temperature. For the investigated parameters, the inherent consistency check, namely the value of the renormalized vertex, does not indicate a failure of the method. Furthermore, the effect proved to be numerically stable and could be traced back to the behavior of the off-diagonal (in Nambu space) component of the retarded self-energy as a function of the superconducting phase difference between the leads. We hope that this result will stimulate further theoretical (and ultimately experimental) research in this regime.

\begin{acknowledgments}
We are grateful to Sabine Andergassen, Luca Dell'Anna, Reinhold Egger, Christoph Karrasch and Matti Laakso for helpful discussions. This work was supported by the DFG-Forschergruppe 723.
\end{acknowledgments}

\appendix
\section{Fourier transformations}
\label{sec:FT}
The goal of the FTs discussed here is to exploit the global periodicity $t_i^{(\prime)}\to t_i^{(\prime)}+T$ with $T=2\pi/V$. For single-particle quantities, the diFT is a valid way of doing so ($\nu \in [-V/2,V/2)$, $n,n^\prime \in \mathbb{Z}$):\cite{Del08,Arn87,Mar99}
\begin{align} F(t_1,t_1^\prime)&=\sum_{n,n^\prime=-\infty}^{\infty} \int_{-V/2}^{V/2} \frac{d \nu}{2 \pi} e^{-i \nu_n t_1 + i \nu_{n^\prime} t_1^\prime} F(\nu)_{n n^\prime} \\ F(\nu)_{n n^\prime} &= \frac{V}{2 \pi} \int_{-\pi /V}^{\pi /V} dt_1^\prime \int_{-\infty}^{\infty} dt_1 e^{i \nu_n t_1 - i \nu_{n^\prime}t_1^\prime} F(t_1,t_1^\prime) \end{align}
The advantage is that contracting quantities corresponds to contracting the discrete indices (i.e. a matrix-matrix-multiplication) while the continuous frequencies are the same on both quantities. This implies that inverting a quantity corresponds to a simple matrix-inversion. The disadvantage is that the diFT cannot be generalized to the two-particle case without losing the property of exploiting the periodicity.

The siFT solves this problem by introducing one centered time and additional relative times. The periodicity is always exploited via the centered time which corresponds to one discrete index while the relative times correspond to real frequencies. The following relation holds in the single-particle case:
\begin{equation}t=\frac{t_1+t_1^\prime}{2} \hspace{1cm} \tau=\frac{t_1-t_1^\prime}{2}\end{equation}
\begin{align} F(t,\tau)&=\frac{1}{2} \sum_{m=-\infty}^{\infty} \int_{-\infty}^{\infty} \frac{d \Omega}{2 \pi} e^{-i \Omega \tau} e^{-i m V t} F(\Omega)_{m} \\ F(\Omega)_{m} &= 2 \frac{V}{2 \pi} \int_{-\pi /V}^{\pi /V} dt \int_{-\infty}^{\infty} d\tau e^{i \Omega \tau} e^{i m V t} F(t,\tau) \end{align}
Note that $\int dt_1 \int d t_1^\prime = 2 \int dt \int d\tau$. For the two-particle case a possible choice is:
\begin{align} \tau_\Pi & =\hphantom{-}(t_1^\prime+t_2^\prime-t_1-t_2)/4
\\ \tau_X & =(-t_1^\prime+t_2^\prime+t_1-t_2)/4
\\ \tau_\Delta & =\hphantom{-}(t_1^\prime-t_2^\prime+t_1-t_2)/4
\\ \tau_P & =\hphantom{-}(t_1^\prime+t_2^\prime+t_1+t_2)/4
\end{align}
\begin{align}
F & (\Pi,X,\Delta)_m 
\\ \notag & =16 \int d\tau_\Pi \int d \tau_X \int d\tau_\Delta \frac{V}{2 \pi} \int_{-\pi/V}^{\pi/V} d\tau_P 
\\ \notag & \hphantom{=16} \times e^{i(\Pi \tau_\Pi + X \tau_X + \Delta \tau_\Delta)} e^{i m V \tau_P} F(\tau_\Pi, \tau_X, \tau_\Delta, \tau_P)
\end{align}
\begin{align}
F & (\tau_\Pi, \tau_X, \tau_\Delta, \tau_P)
\\ \notag & = \frac{1}{16} \sum_{m} \int \frac{d \Pi}{2 \pi} \int \frac{d X}{2 \pi} \int \frac{d \Delta}{2 \pi} \\ \notag & \hphantom{=16} \times e^{-i(\Pi \tau_\Pi + X \tau_X + \Delta \tau_\Delta)} e^{-i m V \tau_P} F(\Pi,X,\Delta)_m
\end{align}
The major drawback of the siFT is that contracting quantities corresponds to rather complicated contraction rules---this also implies that inverting a quantity in this picture is not possible. For example, the contraction
\begin{equation}
f(\omega_1,\omega_3) =\int \frac{d \omega_2}{2 \pi} f^1(\omega_1,\omega_2) f^2(\omega_2,\omega_3)
\end{equation}
in regular Fourier space corresponds to the following one in siFT:
\begin{align}
f_\textrm{siFT}(\Omega)_m =\sum_{m_1} & f_\textrm{siFT}^1(\Omega+[m-m_1]V)_{m_1}
\\ \notag & \times f_\textrm{siFT}^2(\Omega-m_1 V)_{m-m_1}
\end{align}
A more complicated example is the following one:
\begin{align}
\notag f(\omega_1,\omega_3,\omega_4,\omega_5) =\int \frac{d \omega_2}{2 \pi} f^1(\omega_1,\omega_2)
f^2(\omega_2,\omega_3,\omega_4,\omega_5)
\end{align}
\begin{align}
\Rightarrow & f_\textrm{siFT} \left(\Pi,X,\Delta\right)_m 
\\ \notag &=\sum_{m_1} f_\textrm{siFT}^1\left(\frac{1}{2}\left(\Pi-X+\Delta\right)+\left(\frac{m}{2}-m_1\right)V\right)_{m_1}
\\ \notag & \hphantom{= \sum_{m_1}} \times f_\textrm{siFT}^2\left(\Pi-m_1 V,X+m_1 V,\Delta-m_1 V\right)_{m-m_1}
\end{align}
Deriving such contraction rules one by one and combining them yields the frequency structure of the pp, e-ph and d-ph channel (see App.\ \ref{sec:Details_fRG}) in the siFT formulation. Parameterizing $\gamma(\Pi,X,\Delta)^\Lambda_m$ as $\gamma_\Lambda$ (or as $\gamma_m^\Lambda$), i.e. setting the external frequencies $\Pi,X,\Delta$ in Eq.\ \ref{eq:Gen_flow_eq_2} to zero, yields the starting point for the \textSigma P2O (or the \textgamma P2O) approximation (note that indices $\alpha$ and $q$ were omitted in this section).

As a last remark, there is a relation between siFT and diFT in the single-particle case:
\begin{align}
F_\textrm{diFT}(\omega)_{n n^\prime}&=F_\textrm{siFT}(\Omega=2\omega+[n+n^\prime]V)_{n-n^\prime}
\\ F_\textrm{siFT}(\Omega)_m&=F_\textrm{diFT}\left(\omega=\frac{\Omega-(2n-m)V}{2}\right)_{n,n-m}
\end{align}
In the second line, $n \in \mathbb{Z}$ must be determined such that $[\Omega-(2n-m)V]/2 \in [-V/2,V/2)$.

\section{Details on general functional RG}
\label{sec:Details_fRG}

The general idea of Keldysh functional renormalization group is portrayed in Refs.\ \onlinecite{Gez07} and \onlinecite{Jak07}. Also, a formal derivation of the flow equations using a generating functional approach can be found there.

In Ref.\ \onlinecite{Jak10a}, the general flow equations are given in a notation that is more compliant with the notation used here. The two lowest-order flow equations are ($x,y,a,b$ denote multi-indices consisting of time, state and Keldysh index; double occurence implies summation/integration over the respective sub-indices):
\begin{equation}\frac{\partial}{\partial \Lambda} \Sigma_{x^\prime | x}^\Lambda= - i \gamma_{x^\prime a^\prime | x a}^\Lambda S_{a | a^\prime}^\Lambda \end{equation}
\begin{align} \label{eq:Gen_flow_eq_2}
\frac{\partial}{\partial \Lambda} & \gamma^{\Lambda}_{x^\prime y^\prime | x y} 
\\ \notag &= - i  \gamma^{\Lambda}_{x^\prime y^\prime a^\prime | x y a} S^\Lambda_{a | a^\prime} + i \gamma^\Lambda_{x^\prime y^\prime | a b}\ S^\Lambda_{a | a^\prime} G^\Lambda_{b | b^\prime} \gamma^\Lambda_{a^\prime b^\prime | x y}
\\ \notag & \hphantom{=} + i \gamma^\Lambda_{x^\prime b^\prime | a y} \left[S^\Lambda_{a | a^\prime} G^\Lambda_{b | b^\prime} + S^\Lambda_{b | b^\prime} G^\Lambda_{a | a^\prime}\right] \gamma^\Lambda_{a^\prime y^\prime | x b}
\\ \notag & \hphantom{=} - i \gamma^\Lambda_{y^\prime b^\prime | y a} \left[S^\Lambda_{a | a^\prime} G^\Lambda_{b | b^\prime} + S^\Lambda_{b | b^\prime} G^\Lambda_{a | a^\prime}\right] \gamma^\Lambda_{a^\prime x^\prime | b x} \end{align}
The single-scale propagator $S$ is defined as $S^\Lambda=-G^\Lambda (\partial G_\textrm{free}^{-1}/\partial \Lambda) G^\Lambda$. As to how the $\Lambda$-dependency is introduced, there are many possibilities. Some are discussed in Ref.\ \onlinecite{Jak10b} (for systems coupled to metallic leads). The hybridization method (roughly described in section \ref{sec:flow_eq}) has a clear physical meaning and has been found to be the most suitable one in the case of metallic leads.\cite{Jak10a}

\begin{figure}
\includegraphics[width=0.45\textwidth]{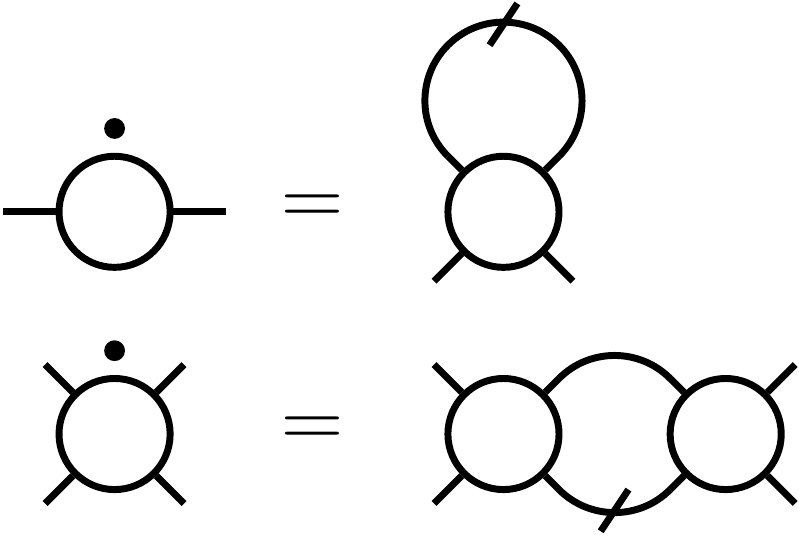}
\caption{\label{fig:07} This plot shows a diagrammatical representation of the flow equations truncated at second order (with $\gamma^{\Lambda=\infty}_{x^\prime y^\prime a^\prime | x y a} =0$). The dot denotes a derivative with respect to $\Lambda$. The lines are full propagators, the crossed lines are single-scale propagators. The circles represent one- or two-particle vertices.}
\end{figure}

The terms on the right-hand-side of the second equation containing the two-particle vertex twice are called particle-particle (pp), exchange particle-hole (e-ph) and direct particle-hole (d-ph) respectively. In a second (or first) order truncation, one sets $\gamma^{\Lambda}_{x^\prime y^\prime a^\prime | x y a} =\gamma^{\Lambda=\infty}_{x^\prime y^\prime a^\prime | x y a} $ (or $\gamma_{x^\prime a^\prime | x a}^\Lambda=\gamma_{x^\prime a^\prime | x a}^{\Lambda=\infty}$---making the second equation redundant). As an illustration, a diagrammatical representation of the flow equations in second order truncation is given in Fig.\ \ref{fig:07}.

\bibliographystyle{apsrev}
\bibliography{paper}

\end{document}